\documentclass[12pt]{article}

\usepackage{epsf}
\usepackage{latexsym,amssymb,euscript}
\usepackage[dvips]{graphicx}
\usepackage{amsmath}

\newcommand{\beq}{\begin{equation}}
\newcommand{\eeq}{\end{equation}}
\newcommand{\bea}{\begin{eqnarray}}
\newcommand{\eea}{\end{eqnarray}}

\makeatletter
\def\appendix{\par\clearpage
  \setcounter{section}{0}
  \setcounter{subsection}{0}
  \@addtoreset{equation}{section}
  \def\@sectname{Appendix~}
  \def\theequation{\thesection.\arabic{equation}}
  \def\theequation{\thesection.\arabic{equation}}
  \def\thesection{\Alph{section}}}
\makeatother

\begin{document}
\begin{titlepage}

\begin{center}
{\LARGE \bf The next-to-leading order forward jet vertex in the
small-cone approximation}
\end{center}

\vskip 0.5cm

\centerline{D.Yu. Ivanov$^{2\P}$ and A.~Papa$^{1\ddagger}$}

\vskip .6cm

\centerline{${}^1$ {\sl Dipartimento di Fisica, Universit\`a della Calabria,}}
\centerline{\sl and Istituto Nazionale di Fisica Nucleare, Gruppo collegato di
Cosenza,}
\centerline{\sl Arcavacata di Rende, I-87036 Cosenza, Italy}

\vskip .2cm

\centerline{${}^2$ {\sl Sobolev Institute of Mathematics and
Novosibirsk State University,}}
\centerline{\sl 630090 Novosibirsk, Russia}

\vskip 2cm

\begin{abstract}
We consider within QCD collinear factorization the process $p+p\to {\rm jet}
+{\rm jet} +X$, where two forward high-$p_T$ jets are produced with a
large separation in rapidity $\Delta y$ (Mueller-Navelet jets).
In this case the (calculable) hard part of the reaction receives large
higher-order corrections  $\sim \alpha^n_s (\Delta y)^n$, which can be
accounted for in the BFKL approach. In particular, we calculate in the
next-to-leading order the impact factor (vertex) for the production of a
forward high-$p_T$ jet, in the approximation of small aperture of the jet
cone in the pseudorapidity-azimuthal angle plane. The final expression for
the vertex turns out to be simple and easy to implement in numerical
calculations.
\end{abstract}


$
\begin{array}{ll}
^{\P}\mbox{{\it e-mail address:}} &
\mbox{d-ivanov@math.nsc.ru}\\
^{\ddagger}\mbox{{\it e-mail address:}} &
\mbox{papa@cs.infn.it}\\
\end{array}
$

\end{titlepage}

\vfill \eject

\section{Introduction}

The production of two forward high-$p_T$ jets in the fragmentation region of
two colliding hadrons at high energies, the so called Mueller-Navelet
jets~\cite{Mueller:1986ey}, is considered an important process for the
manifestation of the BFKL~\cite{BFKL} dynamics at hadron colliders, such as
Tevatron and LHC.

The theoretical investigation of this process implies a combined use
of collinear and BFKL factorization: the process is started by two hadrons
each emitting one parton, according to its parton distribution function
(PDF), which obeys the standard DGLAP evolution~\cite{DGLAP}. On the other
side, at large squared center of mass energy $\sqrt{s}$, i.e. when the
rapidity gap between the two produced jets is large, the BFKL resummation
comes into play, since large logarithms of the energy compensate the
small QCD coupling and must be resummed to all orders of perturbation
theory.

The BFKL approach provides a general framework for this resummation in
the leading logarithmic approximation (LLA), which means resummation of all
terms $(\alpha_s\ln(s))^n$, and in the next-to-leading logarithmic
approximation (NLA), which means resummation of all terms
$\alpha_s(\alpha_s\ln(s))^n$. Such resummation is process-independent and
is encoded in the Green's function for the interaction of two
Reggeized gluons. The Green's function is determined through the BFKL
equation, which is an iterative integral equation, whose kernel is known at
the next-to-leading order (NLO) both for forward scattering (i.e. for $t=0$
and color singlet in the $t$-channel)~\cite{FL98,CC98} and for any fixed
(not growing with energy) momentum transfer $t$ and any possible two-gluon
color state in the $t$-channel~\cite{FF05}.

The process-dependent part of the information needed for constructing the
cross section for the production of Mueller-Navelet jets is contained
in the impact factors for the transition from the colliding
parton to the forward jet (the so called ``jet vertex'').

Such impact factors were calculated with NLO accuracy in~\cite{Bartels:2002yj},
where a careful analysis was performed, based on the separation of the
various rapidity regions and on the isolation of the collinear divergences
to be adsorbed in the renormalization of the PDFs. The results
of~\cite{Bartels:2002yj} were then used in~\cite{Colferai:2010wu} for a
numerical estimation in the NLA of the cross section for Mueller-Navelet
jets at LHC and for the analysis of the azimuthal correlation of the produced
jets. This numerical analysis followed previous
ones~\cite{Vera:2007kn,Marquet:2007xx} based on the inclusion of NLO effects
only in the Green's functions. Recently we performed a new
calculation~\cite{Caporale:2011cc} of the jet impact factor, confirming the
results of ~\cite{Bartels:2002yj}.

In this paper we recalculate the NLO impact factor for the production of
forward jets in the ``small-cone'' approximation
(SCA)~\cite{Furman:1981kf,Aversa}, i.e. for small jet cone aperture in the
rapidity-azimuthal angle plane. Our starting point are the totally inclusive
NLO parton impact factors calculated in~\cite{FFKP99}, according to the general
definition in the BFKL approach given in Ref.~\cite{FF98}. The calculation is
lengthy, but straightforward, since the standard BFKL definition of impact
factor provides the route to be followed. The use of the SCA, moreover, allows
to get a simple analytic result for the jet vertices, easily implementable in
numerical calculations  and therefore particularly suitable for a
semi-analytical cross-check of the numerical approaches which treat the cone
size exactly.

The paper is organized as follows. In the next Section we will present the
factorization structure of the cross section, recall the definition
of BFKL impact factor and discuss the treatment of the divergences arising
in the calculation; in Section~3 we describe the procedure for the jet
definition and the SCA; in Section~4 and~5 we present the details of the
calculation at LO and NLO, respectively; in Section~6 we draw some conclusions.

\begin{figure}[tb]
\centering
\includegraphics{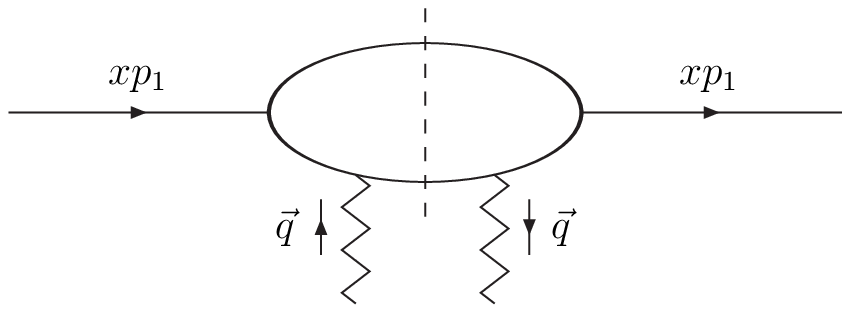}
\caption{Diagrammatic representation of the forward parton impact factor.}
\label{fig:if}
\end{figure}

\section{General framework}

We consider the process
\beq
p(p_1)+p(p_2)\to {\rm jet}(k_1)+{\rm jet}(k_2)+X \
\label{process}
\eeq
in the kinematical region where the jets have large transverse
momenta\footnote{See Eq.~(\ref{sudakov}) below for the definition
of the transverse part of a 4-vector. },
$\vec k_{1}^{\,2}\sim \vec k_{2}^{\,2} \gg \Lambda_{\rm QCD}^2$.
This provides the hard scale, $Q^2\sim \vec k_{1,2}^{\,2}$, which makes
perturbative QCD methods applicable. Moreover, the energy of the proton
collision is assumed to be much bigger than the hard scale,
$s=2p_1\cdot p_2\gg \vec k_{1,2}^{\,2}$.

We consider the leading behavior in the $1/Q$-expansion (leading twist
approximation). With this accuracy one can neglect the masses of initial
protons. The state of the jets can be described completely by their
(pseudo)rapidities\footnote{For
massless particle the rapidity coincides with pseudorapidity, $y=\eta$,
the latter being related to the particle polar scattering angle by
$\eta=-\ln\tan \frac{\theta}{2}$.} $y_{1,2}$ and transverse momenta
$\vec k_{1,2}$.
Moreover, we denote the azimuthal angles of the jets as $\phi_{1,2}$.

\begin{figure}[tb]
\centering
\includegraphics{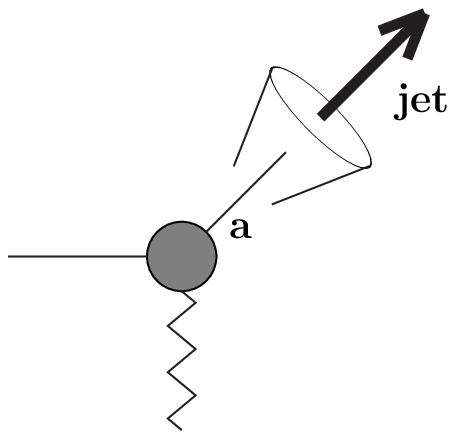}
\caption{Parton-Reggeon collision, the jet is formed by a single parton.}
\label{fig:if-lo}
\end{figure}

In QCD collinear factorization the cross section of the process reads
\beq
\frac{d\sigma}{dy_1dy_2d^2\vec k_1d^2\vec k_2} =\sum_{i,j=q,g}\int\limits^1_0
\int\limits^1_0 dx_1dx_2 f_i(x_1,\mu_F) f_j(x_2,\mu_F)
\frac{d\hat \sigma(x_1 x_2 s,\mu_F)}{dy_1dy_2d^2\vec k_1d^2\vec k_2}\;,
\label{ff}
\eeq
where the $i,j$ indices specify parton types, $i,j=q,\bar q, g$, $f_i(x,\mu_F)$
are the proton PDFs, the longitudinal fractions of the partons involved in
the hard subprocess are $x_{1,2}$, $\mu_F$ is the factorization scale and
$d\hat \sigma(x_1 x_2 s,\mu_F)$ is the partonic cross section for the jet
production.

It is convenient to define the Sudakov decomposition for the jet momenta,
\beq
k_1= \alpha_1 p_1+ \frac{\vec k^{\,2}_1}{\alpha_1 s}p_2+k_{1, \, \perp} \ ,
\quad
k_{1,\, \perp}^2=-\vec k_1^{\,2} \ ,
\label{sudakov}
\eeq
\[
k_2= \frac{\vec k^{\,2}_2}{\alpha_2 s}p_1+\alpha_2 p_2 +k_{2, \, \perp} \ ,
\quad
k_{2,\, \perp}^2=-\vec k_2^{\,2} \ ,
\]
where the jet longitudinal fractions $\alpha_{1,2}$ are related to the jet
rapidities by
\beq
y_1=\frac{1}{2}\ln\frac{\alpha^2_1 s}{\vec k^{\,2}_1}\ , \quad
y_2=-\frac{1}{2}\ln\frac{\alpha^2_2 s}{\vec k^{\,2}_2} \, ,
\label{rapidities}
\eeq
and $dy_1=\frac{d\alpha_1}{\alpha_1}$, $dy_2=-\frac{d\alpha_2}{\alpha_2}$ in
the center of mass system.

We consider the kinematics when the interval of rapidity between the two jets,
\beq
\Delta y=y_1-y_2=\ln\frac{\alpha_1\alpha_2 s}{|\vec k_1||\vec k_2|}\ ,
\label{rap-int}
\eeq
is large. Since the jet longitudinal fractions are equal or smaller (in the
case of additional QCD radiation) than the ones of the participating partons,
$\alpha_1\leq x_1$, $\alpha_2\leq x_2$, we are in a situation where
the energy of the partonic subprocess is much larger than jet transverse
momenta, $x_1x_2 s\gg \vec k^{\,2}_{1,2}$ ($\vec k^{\,2}_{1}$ and
$\vec k^{\,2}_{2}$ are considered to be of similar order $\sim \vec k^{\,2}$).
In this region the perturbative partonic cross section receives at higher
orders large contributions $\sim \alpha^n_s\ln^n \frac{s}{\vec k^{\,2}}$,
related with large energy logarithms. It is the aim of this paper to
elaborate the resummation of such enhanced contributions with NLA accuracy
using the BFKL approach.

Let us remind some generalities of the BFKL method. Due to the optical theorem,
the cross section is related to the imaginary part of the forward
proton-proton scattering amplitude,
\beq
\sigma =\frac{{\cal I}m_s A}{s} \ .
\eeq
In the BFKL approach the kinematic limit $s\gg \vec k^{\,2}$
 of the
forward amplitude may be presented in $D$ dimensions as follows:
\beq
{\cal I}m_s
\left( {\cal A} \right)=\frac{s}{(2\pi)^{D-2}}\int\frac{d^{D-2}\vec q_1}{\vec
q_1^{\,\, 2}}\Phi_1(\vec q_1,s_0)\int
\frac{d^{D-2}\vec q_2}{\vec q_2^{\,\,2}} \Phi_2(-\vec q_2,s_0)
\int\limits^{\delta +i\infty}_{\delta
-i\infty}\frac{d\omega}{2\pi i}\left(\frac{s}{s_0}\right)^\omega
G_\omega (\vec q_1, \vec q_2)\, ,
\label{bfkl-ampl}
\eeq
where the Green's function obeys the BFKL equation
\beq
\omega \, G_\omega (\vec q_1,\vec q_2)  =\delta^{D-2} (\vec q_1-\vec q_2)
+\int d^{D-2}\vec q \, K(\vec q_1,\vec q) \,G_\omega (\vec q, \vec q_1) \;.
\eeq
What remains to be calculated are the NLO impact factors $\Phi_1$ and
$\Phi_2$ which describe the inclusive production of the two jets,
with fixed transverse momenta $\vec k_1$, $\vec k_2$ and
rapidities $y_1$, $y_2$, in the fragmentation regions of the colliding protons
with momenta $p_1$ and $p_2$, respectively.
The energy scale parameter $s_0$ is arbitrary, the amplitude, indeed,
does not depend on its choice within NLA accuracy due to the properties of
NLO impact factors to be discussed below.

For definiteness, we will consider
the case when the jet belongs to the fragmentation region of the proton with
momentum $p_1$, i.e. the jet is produced in the collision of the proton with
momentum $p_1$ off a Reggeon with incoming (transverse) momentum $q$
and denote for shortness in what follows its transverse momentum and
longitudinal fraction by $\vec k$ and $\alpha$, respectively.

Technically, this is done using as starting point the definition of
inclusive parton impact factor, given in Ref.~\cite{FFKP99}, for the
cases of incoming quark(antiquark) and gluon, respectively (see Fig.~1).
Here we review the important steps and give the formulae for the LO parton
impact factors.

Note that both the kernel of the equation for the BFKL Green's function and
the parton impact factors can be expressed in terms of the gluon Regge
trajectory,
\begin{equation}
j(t)\;=\;1\:+\:\omega (t)\;,
\end{equation}
and the effective vertices for the Reggeon-parton interaction.

To be more specific, we will give below the formulae for the case of forward
quark impact factor considered in $D=4+2\epsilon$ dimensions of dimensional
regularization.
We start with the LO, where the quark impact factors are given by
\begin{equation}
\Phi _{q}^{(0)}(\vec q\,)\;=\;\sum_{\{a\}}\:\int \:\frac{dM_a^2}{2\pi }
\:\Gamma_{a q}^{(0)}(\vec q\,)\:[\Gamma_{a q}^{(0)}(\vec q\,)]^*
\:d\rho_a\;,
\label{eq:a7}
\end{equation}
where $\vec q$ is the Reggeon transverse momentum, and $\Gamma ^{(0)}_{a q}$
denotes the Reggeon-quark vertices in the LO or Born approximation.
The sum $\{a\}$ is over all intermediate states $a$ which contribute to the
$q\rightarrow q$ transition. The phase space element $d\rho_a$ of a state $a$,
consisting of particles with momenta $\ell_n$, is ($p_q$ is initial quark
momentum)
\begin{equation}
d\rho _a\;=\;(2\pi )^D\:\delta^{(D)} \left( p_q+q-\sum_{n\in a}\ell
_n\right) \:\prod_{n\in a}\:\frac{d^{D-1}\ell _n}{(2\pi )^{D-1}2E_n}\;,
\label{eq:a8}
\end{equation}
while the remaining integration in (\ref{eq:a7}) is over the squared
invariant mass of the state $a$,
\[
M_a^2\;=\;(p_q+q)^2\; .
\]

In the LO the only intermediate state which contributes is a one-quark state,
$\{a\}=q$. The integration in Eq.~(\ref{eq:a7}) with the known Reggeon-quark
vertices $\Gamma _{q q}^{(0)}$ is trivial and the quark impact factor reads
\begin{equation}
\Phi _{q}^{(0)}(\vec q\,
)\;=g^2 \frac{\sqrt{N^2-1}}{2N} \;,
\label{eq:a77}
\end{equation}
where $g$ is QCD coupling, $\alpha_s=g^2/(4\pi)$, $N=3$ is the number of QCD
colors.

In the NLO the expression~(\ref{eq:a7}) for the quark impact factor has to be
changed in two ways. First one has to take into account the radiative
corrections to the vertices,
$$
\Gamma _{q q}^{(0)}\to \Gamma_{q q}=\Gamma_{q q}^{(0)}+\Gamma _{q q}^{(1)} \;.
$$
Secondly, in the sum over $\{a\}$ in~(\ref{eq:a7}), we have to include more
complicated states which appear in the next order of perturbative theory. For
the quark impact factor this is a state with an additional gluon, $a=q g$.
However, the integral over $M_a^2$ becomes divergent when an extra gluon
appears in the final state. The divergence arises because the gluon may be
emitted not only in the  fragmentation region of initial quark, but also in
the central rapidity region. The contribution of the central region must be
subtracted from the impact factor, since it is to be assigned in the BFKL
approach to the Green's function. Therefore the result for the forward quark
impact factor reads
\begin{eqnarray}
\label{eq:a19}
\Phi _{q}(\vec q\, ,s_0)
=\left(\frac{ s_0}{\vec q^{\: 2} }\right)^{\omega(-\vec q^{\: 2})}
\:\sum_{\{a\}}\:\int \:\frac{dM_a^2}{2\pi }\:\Gamma
_{aq}(\vec q\,)\:[\Gamma _{aq}(\vec q\,)]^*\:d\rho _a\:\theta(s_\Lambda -M_a^2)
&&
\nonumber \\
-\frac{1}{2}\int d^{D-2} k\frac{\vec q^{\,\, 2}}{\vec k^{\, 2}}
\Phi_{q}^{(0)}(\vec k)\mathcal{K}_r^{(0)}(\vec k,\vec q\,)
\ln\left( \frac{s_{\Lambda}^2}{(\vec k-\vec q\,)^2 s_0}\right) \;.
&&
\end{eqnarray}
The second term in the r.h.s. of Eq.~(\ref{eq:a19}) is the subtraction
of the gluon emission in the central rapidity region. Note that, after this
subtraction, the intermediate parameter $s_\Lambda $ in the r.h.s. of
Eq.~(\ref{eq:a19}) should be sent to infinity.
The dependence on $s_\Lambda$ vanishes because of the cancellation between the
first and second terms. $K_r^{(0)}$ is the part of LO BFKL kernel related to
real gluon production,
\begin{equation}
\label{eq:a20}
K_r^{(0)}(\vec k,\vec q\,)\;=\;\frac{2 g^2N}{(2\pi )^{D-1}}
\frac{1}{(\vec k-\vec q\,)^2} \; .
\end{equation}
The factor in Eq.~(\ref{eq:a19}) which involves the Regge trajectory arises
from the change of energy scale~($\vec q^{\: 2}\to s_0$) in the vertices
$\Gamma$. The trajectory function $\omega (t)$ can  be taken here in the
one-loop approximation ($t=-\vec q^{\,\,2}$),
\begin{equation}
\omega (t)\;=\;\frac{g^2t}{(2\pi )^{D-1}}\frac{N}2\int \frac{d^{D-2}k}
{\vec k^2(\vec q-\vec k)^2}\;=\;-\;g^2N\frac{\Gamma (1-\varepsilon )}
{(4\pi )^{D/2}} \frac{\Gamma^2(\varepsilon )}{\Gamma (2\varepsilon )}
(\vec q^{\,\,2})^\varepsilon \; . \label{eq:b20}
\end{equation}

In the Eqs.~(\ref{eq:a7}) and (\ref{eq:a19}) we suppress for shortness the
color indices (for the explicit form of the vertices see~\cite{FFKP99}).
The gluon impact factor $\Phi_g(\vec q\,)$ is defined similarly. In the gluon
case only the single-gluon intermediate state contributes in the LO, $a=g$,
which results in
\begin{equation}
\Phi _{g}^{(0)}(\vec q\,
)\;=\frac{C_A}{C_F}\Phi _{q}^{(0)}(\vec q\,) \; ,
\label{eq:a77a}
\end{equation}
here $C_A=N$ and $C_F=(N^2-1)/(2N)$. Whereas in NLO additional two-gluon,
$a=g g$, and quark-antiquark, $a=q \bar q$, intermediate states have to be
taken into account in the calculation of the gluon impact factor.

The definition of inclusive parton impact factors involves the integration
over all possible intermediate states appearing in the parton-Reggeon
collision. Up to the next-to-leading order, this means that we can have one
or two partons in the intermediate state. Then, in order to allow for the
inclusive production of a jet, these integrations must be suitably constrained
to take into account that the kinematics of the parton or the pair of partons
which generate the jet is fixed by the jet kinematics.

\section{Jet definition and small-cone approximation}

\begin{figure}[tb]
\centering
\includegraphics{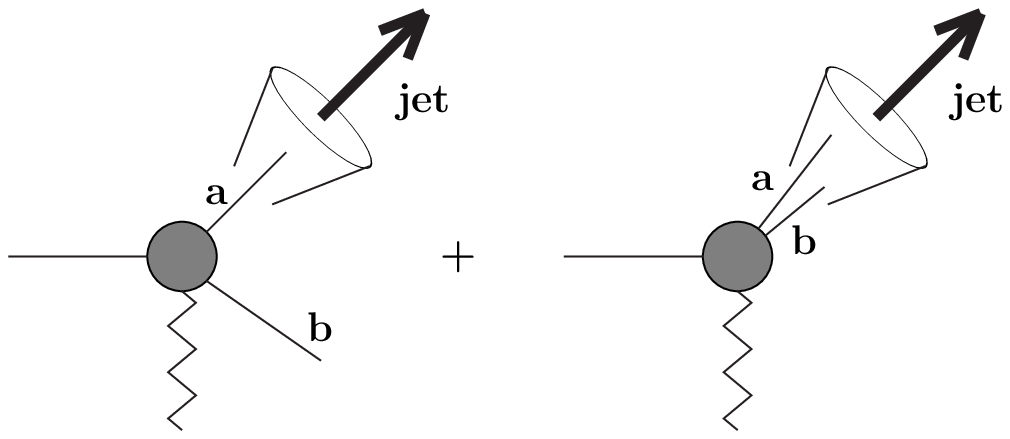}
\caption{Parton-Reggeon collision, two partons are produced and the jet is
formed either by one of the partons or by both partons.}
\label{fig:if-nlo}
\end{figure}

At LO the (totally inclusive) parton impact factor takes contribution
from a one-particle intermediate state; equivalently, only one parton is
produced in the collision between the incoming parton and the Reggeon,
as shown in Fig.~\ref{fig:if-lo}.
Therefore, the kinematics of the produced parton is totally fixed by the
jet kinematics. At NLO we have both the virtual corrections (which have the
kinematical structure shown in Fig.~\ref{fig:if-lo}) and also two-particle
production in the parton-Reggeon collision. The jet in the latter case can be
either produced by one of the two partons or by both together. If we call the
produced partons $a$ and $b$, we have the following contributions, as shown
in Fig.~\ref{fig:if-nlo} (see, for instance, Ref.~\cite{Jager:2004jh}):
\begin{enumerate}
\item the parton $a$ generates the jet, while the parton $b$ can have arbitrary
kinematics, provided that it lies {\em outside} the jet cone;

\item similarly with $a\leftrightarrow b$;

\item the two partons $a$ and $b$ both generate the jet.
\end{enumerate}

\begin{figure}[tb]
\centering
\includegraphics{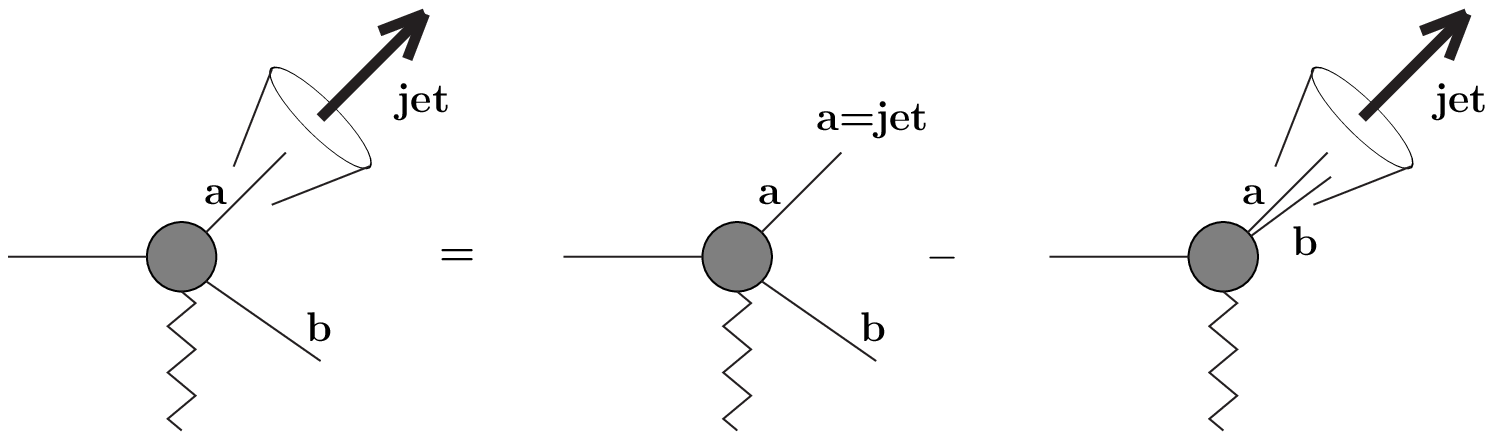}
\caption{The production of the jet by one parton when the second one is
outside the cone can be seen as the ``inclusive'' production minus the
contribution when the second parton is inside the cone.}
\label{fig:if-nlo-incl}
\end{figure}

The cases 1. and 2. are replaced  in the actual calculation by the following
two (as illustrated in Fig.~\ref{fig:if-nlo-incl}):
\begin{enumerate}
\item the parton $a$ generates the jet, while the parton $b$ can have arbitrary
kinematics (``inclusive'' jet production by the parton $a$); then, the case
when the parton $b$ lies {\em inside} the jet cone is subtracted;

\item similarly with $a\leftrightarrow b$.
\end{enumerate}

Let us introduce now the ``small-cone'' approximation (SCA). In view of
the discussion above, we should define it in the two cases of jet generated by
one parton or by two partons.

The relative rapidity and azimuthal angle between the two partons are
\[
\Delta y=\frac{1}{2}\ln\frac{\zeta^2 (\vec k-\vec q)^2}{\bar \zeta^2 \vec k^{\,2}}
\;, \;\;\;\;\;
\Delta \phi=\arccos \frac{\vec q\cdot\vec k-\vec k^{\,2}}{|\vec k||\vec q-\vec k|}
\;,
\;\;\;\;\;
\bar \zeta \equiv 1-\zeta\;.
\]

Let the parton with momentum $\vec k$ and longitudinal fraction $\zeta$
generate the jet, whereas the other parton (with momentum $\vec q-\vec k$ and
longitudinal fraction $\bar \zeta$) is a spectator. We introduce the
vector $\vec \Delta$ such that
\[
\vec q=\frac{\vec k}{\zeta}+\vec \Delta\;.
\]
Then, for $\vec\Delta\to 0$ we have
\[
\Delta \phi^2=\frac{\zeta^2}{\bar \zeta^2}\left(\frac{\vec\Delta^2}{\vec k^{\,2}}
-\frac{(\vec k\cdot\vec \Delta)^2}{\vec k^{\,4}}\right)\;, \;\;\;\;\;
\Delta y=\frac{\zeta}{\bar \zeta}\frac{(\vec k\cdot\vec \Delta)}{\vec k^{\,2}}\;,
\]
thus the condition of cone with aperture smaller than $R$ in the
rapidity-azimuthal angle plane becomes
\[
\Delta \phi^2+\Delta y^2=\frac{\zeta^2}{\bar \zeta^2}\frac{\vec\Delta^2}{\vec
k^2} \leq R^2
\]
and therefore
\[
|\vec \Delta|\leq \frac{\bar \zeta}{\zeta} |\vec k| R\;.
\]

The situation is different when both partons form a jet. In this case the
jet momentum is $\vec k=\vec k_1+\vec k_2$ and the jet fraction is
$1=\zeta+\bar \zeta$. The relative rapidity and azimuthal angle between the
jet and the first (second) parton are
\[
\Delta y_1=\frac{1}{2}\ln\frac{\vec k_1^{\,2}}{ \zeta^2 \vec k^{\,2}}\;, \;\;\;\;\;
\Delta \phi_1=\arccos \frac{\vec k\cdot\vec k_1}{|\vec k_1| |\vec k|}\;,
\]
\[
\Delta y_2=\frac{1}{2}\ln\frac{ (\vec k_1-\vec k)^2}{\bar \zeta^2  \vec k^{\,2}}\;,
\;\;\;\;\;
\Delta\phi_2=\arccos\frac{\vec k\cdot(\vec k-\vec k_1)}{|\vec k|
|\vec k-\vec k_1|}\;.
\]
Introducing now the vector $\vec \Delta$ as
\[
\vec k_1=\zeta \vec k+\vec \Delta \;,
\]
we find
\[
\Delta y_1^2+\Delta \phi_1^2=\frac{\vec\Delta^2}{\zeta^2 \vec k^{\,2}}\;,\;\;\;\;\;
\Delta y_2^2+\Delta \phi_2^2=\frac{\vec\Delta^2}{\bar \zeta^2 \vec k^{\,2}}\;,
\]
so that the requirement that both partons are inside the cone is now
\[
|\vec \Delta|\leq R\, |\vec k| \, \min(\zeta,\bar \zeta)\;.
\]

\section{Impact Factor in the LO}


The inclusive LO impact factor of proton may be thought of as the convolution
of quark and gluon impact factors,
given in Eqs.~(\ref{eq:a77},\ref{eq:a77a}),
with the corresponding proton PDFs,
\beq
d\Phi={\cal C} \, dx \left(\frac{C_A}{C_F}f_g(x)+\sum_{a=q,\bar q} f_a(x)
\right)\ , \quad {\cal C}=g^2\frac{\sqrt{N^2-1}}{2N}=2\pi\alpha_s
\sqrt{\frac{2\, C_F}{C_A}}\;.
\label{inclusive}
\eeq

In order to establish the proper normalization for the jet impact factor,
we insert into the inclusive impact factor~(\ref{inclusive}) the delta
functions which depend on the jet variables, transverse momentum $\vec k$
and longitudinal fraction $\alpha$:
\beq
\frac{d\Phi^J}{\vec q^{\,\, 2}}={\cal C}\int \,d\alpha
\frac{d^2\vec k}{\vec k^{\,2}}\, dx \, \delta^{(2)}\left(\vec k-\vec q\right)
\delta(\alpha-x)\left(\frac{C_A}{C_F}f_g(x)+\sum_{a=q,\bar q} f_a(x)\right)\;.
\eeq

In what follows we will calculate the projection of the impact factor
on the eigenfunctions of LO BFKL kernel, i.e. the impact factor
in the so called $(\nu,n)$-representation,
\beq
\Phi(\nu,n)=\int d^2\vec q \,\frac{\Phi(\vec q)}{\vec q^{\,\, 2}}
\frac{1}{\pi \sqrt{2}} \left(\vec q^{\,\, 2}\right)^{i\nu-\frac{1}{2}}
e^{i n \phi}\;.
\label{nu_rep}
\eeq
Here $\phi$ is the azimuthal angle of the vector $\vec q$ counted from
some fixed direction in the transverse space.

\section{NLO calculation}

We will work in $D=4+2\epsilon$ dimensions and calculate the NLO impact factor
directly in the $(\nu,n)$-representation~(\ref{nu_rep}), working out
separately virtual corrections and real emissions. To this purpose we introduce
the ``continuation'' of the LO BFKL eigenfunctions to non-integer
dimensions,
\beq
\left(\vec q^{\,\, 2} \right)^{\gamma}e^{i n \phi}\to  \left(\vec q^{\,\,2}
\right)^{\gamma-{n \over 2}}\left(\vec q \cdot \vec l \,\, \right)^n\;,
\label{eigen}
\eeq
where $\gamma=i\nu-\frac{1}{2}$ and  $\vec l^{\:2}=0$. It is assumed that the
vector $\vec l$ lies only in the first two of the $2+2\epsilon$ transverse
space dimensions, i.e. $\vec l=\vec e_1+i\,  \vec e_2$, with
$\vec e_{1,2}^{\:2}=1$, $\vec e_{1}\cdot \vec e_2=0$. In the limit
$\epsilon\to 0$ the r.h.s. of Eq.~(\ref{eigen}) reduces to the LO BFKL
eigenfunction.
This technique was used recently in Ref.~\cite{Kirschner:2009qu}. An even
more general method, based on an expansion in traceless products, was uses
earlier in Ref.~\cite{Kotikov:2000pm} for the calculation of NLO BFKL kernel
eigenvalues. In the case of interest, $\vec l^{\:2}=0$, these two approaches
lead, actually, to similar formulas.

Thus, for the case of non-integer dimension the LO result for the impact
factor reads
\beq
\frac{d\Phi^J}{\vec q^{\,\, 2}}={\cal C} \,d\alpha \frac{d^{2+2\epsilon}\vec k}{\vec k^{\,2}} \, \delta^{(2+2\epsilon)}\left(\vec k-\vec q\right)
\left(\frac{C_A}{C_F}f_g(\alpha)+\sum_{a=q,\bar q} f_a(\alpha)\right)\;,
\eeq
which in the $(\nu,n)$-representation gives the result
\beq
\frac{\pi\sqrt{2}\, \vec k^{\,2}}{{\cal C}}\frac{d\Phi^J(\nu,n)}
{d\alpha d^{2+2\epsilon}\vec k}=
\left( \frac{C_A}{C_F}f_g(\alpha)
+\sum_{a=q,\bar q}f_a(\alpha)\right)
\left(\vec k^{\,2} \right)^{\gamma-{n \over 2}}
\left(\vec k \cdot \vec l \,\, \right)^n \;.
\label{LO}
\eeq

Collinear singularities which appear in the NLO calculation are removed by the
renormalization of PDFs. The relations between the bare and renormalized
quantities are
\bea
&
f_q(x)=f_q(x,\mu_F)-\frac{\alpha_s}{2\pi}\left(\frac{1}{\hat \epsilon}
+\ln\frac{\mu_F^2}{\mu^2}\right)
\int\limits^1_x\frac{dz}{z}\left[P_{qq}(z)f_q(\frac{x}{z},\mu_F)
+P_{qg}(z)f_g(\frac{x}{z},\mu_F)\right]\;, &\nonumber\\
&
f_g(x)=f_g(x,\mu_F)-\frac{\alpha_s}{2\pi}\left(\frac{1}{\hat \epsilon}
+\ln\frac{\mu_F^2}{\mu^2}\right)
\int\limits^1_x\frac{dz}{z}\left[P_{gq}(z)f_q(\frac{x}{z},\mu_F)
+P_{gg}(z)f_g(\frac{x}{z},\mu_F)\right]\;,
\label{DGLAPpdfs}
\eea
where $\frac{1}{\hat \epsilon}=\frac{1}{\epsilon}+\gamma_E-\ln (4\pi)\approx
\frac{\Gamma(1-\epsilon)}{\epsilon (4\pi)^\epsilon}$, and the DGLAP kernels
are given by
\bea
P_{gq}(z)&=&C_F\frac{1+(1-z)^2}{z} \;, \\
P_{qg}(z)&=&T_R\left[z^2+(1-z)^2\right]\;, \\
P_{qq}(z)&=&C_F\left( \frac{1+z^2}{1-z} \right)_+
= C_F\left[ \frac{1+z^2}{(1-z)_+} +{3\over 2}\delta(1-z)\right]\;, \\
P_{gg}(z)&=&2C_A\left[\frac{1}{(1-z)_+} +\frac{1}{z} -2+z(1-z)\right]
+  \left({11\over 6}C_A-\frac{n_f}{3}\right)\delta(1-z)\;,
\eea
with $T_R=1/2$. Here and below we always adopt the $\overline{\rm{MS}}$ scheme.

Now we can calculate the collinear counterterms which appear due to the
renormalization of the bare PDFs. Inserting the expressions given in
Eqs.~(\ref{DGLAPpdfs}) into the LO impact factor~(\ref{LO}), we obtain
\beq
\frac{\pi\sqrt{2}\, \vec k^{\,2}}{{\cal C}}
\frac{d\Phi^J(\nu,n)|_{\rm{collinear \ c.t.}}}
{d\alpha d^{2+2\epsilon}\vec k}
= -  \frac{\alpha_s}{2\pi}\left(\frac{1}{\hat \epsilon}+\ln\frac{\mu_F^2}
{\mu^2}\right)
\left(\vec k^{\,2} \right)^{\gamma-{n \over 2}}\left(\vec k \cdot \vec l
\,\, \right)^n \int\limits^1_{\alpha} \frac{dz}{z}
\label{c.count.t}
\eeq
\[
\times\left[ \sum_{a=q,\bar q}\left( P_{qq}(z)f_a\left(\frac{\alpha}{z}\right)
+  P_{qg}(z) f_g\left(\frac{\alpha}{ z}\right) \right)
+\frac{C_A}{C_F}\left( P_{gg}(z) f_g\left(\frac{\alpha}{z}\right)
+  P_{gq}(z)\sum_{a=q,\bar q}f_a\left(\frac{\alpha}{z}\right)\right)
\right]\;.
\]

The other counterterm is related with the QCD charge renormalization,
\beq
\alpha_s=\alpha_s(\mu_R)\left[1+\frac{\alpha_s(\mu_R)}{4\pi}\beta_0
\left(\frac{1}{\hat \epsilon}+\ln\frac{\mu_R^2}{\mu^2}\right)\right]
\,,\quad \beta_0=\frac{11C_A}{3}-\frac{2n_f}{3}\;,
\label{charge-ren}
\eeq
and is given by
\[
\frac{\pi\sqrt{2}\, \vec k^{\,2}}{{\cal C}}\frac{d\Phi^J(\nu,n)|_{\rm{charge
\ c.t.}}} {d\alpha d^{2+2\epsilon}\vec k}
= \frac{\alpha_s}{2\pi}\left(\frac{1}{\hat \epsilon}+\ln\frac{\mu_R^2}
{\mu^2}\right)
\left(\vec k^{\,2} \right)^{\gamma-{n \over 2}}\left(\vec k \cdot \vec l
\,\, \right)^n \int\limits^1_{\alpha} \frac{dz}{z}\,  \delta(1-z)
\]
\beq
\times\left( \sum_{a=q,\bar q} f_a\left(\frac{\alpha}{z}\right)
+\frac{C_A}{C_F}  f_g\left(\frac{\alpha}{z}\right)
\right)\left(\frac{11 C_A}{6}-\frac{n_f}{3}\right) \;.
\label{charge.count.t}
\eeq

To simplify formulae, from now on we put the arbitrary scale of dimensional
regularization equal to the unity, $\mu=1$.

In what follows we will present intermediate results always for
$\frac{\pi\sqrt{2}\, \vec k^{\,2}}{{\cal C}}\frac{d\Phi^J(\nu,n)}
{d\alpha d^{2+2\epsilon}\vec k}$, which we denote  for shortness  as
\beq
\frac{\pi\sqrt{2}\, \vec k^{\,2}}{{\cal C}}\frac{d\Phi^J(\nu,n)}
{d\alpha d^{2+2\epsilon}\vec k}\equiv I \, .
\label{def_I}
\eeq
Moreover, $\alpha_s$ with no argument can always be understood as
$\alpha_s(\mu_R)$.

We will consider separately the subprocesses initiated by a quark and a gluon
PDF, and denote
\beq
I=I_q+I_g \, .
\label{def_I_q+g}
\eeq
We start with the case of incoming quark.

\subsection{Incoming quark}

We distinguish virtual corrections and real emission contributions,
\beq
I_q=I_q^V+I_q^R \; .
\eeq

Virtual corrections are the same as in the case of the inclusive quark impact
factor, therefore we have
\[
I_q^V=-\frac{\alpha_s}{2\pi}\frac{\Gamma[1-\epsilon]}{(4\pi)^\epsilon}
\frac{1}{\epsilon}\frac{\Gamma^2(1+\epsilon)}{\Gamma(1+2\epsilon)}
\left(\vec k^{\,2} \right)^{\gamma+\epsilon-{n \over 2}}
\left(\vec k \cdot \vec l \,\, \right)^n \int\limits^1_{\alpha}\frac{d \zeta}
{\zeta} \, \delta (1-\zeta)\sum_{a=q,\bar q}f_a\left(\frac{\alpha}{\zeta}
\right)
\]
\[
\times\left\{
C_F\left(\frac{2}{\epsilon}-\frac{4}{1+2\epsilon}+1\right)\right.
-n_f\frac{1+\epsilon}{(1+2\epsilon)(3+2\epsilon)}
+C_A\left(\ln\frac{s_0}{\vec k^{\,2}}+\psi(1-\epsilon)-2\psi(\epsilon)
+\psi(1)\right.
\]
\beq
\left.\left.
+\frac{1}{4(1+2\epsilon)(3+2\epsilon)}-\frac{2}{\epsilon(1+2\epsilon)}
-\frac{7}{4(1+2\epsilon)}-\frac{1}{2}\right)\right\}\;.
\label{Qvirt}
\eeq
Note that the contribution $\sim \ln\frac{s_0}{\vec k^{\,2}}$ in
Eq.~(\ref{Qvirt}) originates from the factor $\left(\frac{ s_0}{\vec q^{\: 2}
}\right)^{\omega(-\vec q^{\: 2})}$ in the definition of the NLO impact factor,
see Eq.~({\ref{eq:a19}}), which is accounted for virtual corrections in the
BFKL approach.

We expand~(\ref{Qvirt}) in $\epsilon$ and present the result as a sum of
the singular and the finite parts. The singular contribution reads
\[
\left(I_q^V\right)_{s}=-\frac{\alpha_s}{2\pi}\frac{\Gamma[1-\epsilon]}
{(4\pi)^\epsilon}\frac{1}{\epsilon}\frac{\Gamma^2(1+\epsilon)}
{\Gamma(1+2\epsilon)}\left(\vec k^{\,2}
\right)^{\gamma+\epsilon-{n \over 2}}\left(\vec k \cdot \vec l \,\, \right)^n
\int\limits^1_{\alpha}\frac{d \zeta}{\zeta} \, \delta (1-\zeta)
\sum_{a=q,\bar q}f_a\left(\frac{\alpha}{\zeta}\right)
\]
\beq
\times \left\{C_F\left(\frac{2}{\epsilon}-3\right)-\frac{n_f}{3}
+C_A\left(\ln\frac{s_0}{\vec k^{\,2}}+\frac{11}{6}\right)
\right\}\;,
\label{Qvirt-div}
\eeq
whereas for the regular part we obtain
\[
\left(I_q^V\right)_{r}=-\frac{\alpha_s}{2\pi}
\left(\vec k^{\,2} \right)^{\gamma-{n \over 2}}
\left(\vec k \cdot \vec l \,\, \right)^n
\int\limits^1_{\alpha}\frac{d \zeta}{\zeta} \, \delta (1-\zeta)
\sum_{a=q,\bar q}f_a\left(\frac{\alpha}{\zeta}\right)
\]
\beq
\times \left\{8 C_F+\frac{5n_f}{9}
-C_A\left(\frac{85}{18}+\frac{\pi^2}{2}\right)\right\}\;.
\label{Qvirt-fin}
\eeq
Note that $\left(I_q^V\right)_{s}+\left(I_q^V\right)_{r}$ differs from
$I_q^V$ by terms which are ${\cal O}(\epsilon)$.

\subsubsection{Quark-gluon intermediate state}
\label{subsub_QG}

The starting point here is the quark-gluon intermediate state contribution
to the inclusive quark impact factor,
\beq
\Phi^{\{QG\}}=\Phi_q g^2\vec q^{\,\, 2}\frac{d^{2+2\epsilon} \vec k_1}
{(2\pi)^{3+2\epsilon}}
\frac{d\beta_1}{\beta_1}\frac{[1+\beta_2^2+\epsilon \beta_1^2]}
{\vec k_1^{\,2} \vec k_2^{\,2} (\vec k_2\beta_1-\vec k_1 \beta_2)^2}
\left\{C_F \beta_1^2\vec k_2^{\,2}+C_A\beta_2\left(\vec k_1^{\,2}
-\beta_1\vec k_1\cdot\vec q\right) \right\}\;,
\eeq
where $\beta_1$ and $\beta_2$ are the relative longitudinal momenta
($\beta_1+\beta_2=1$) and $\vec k_1$ and $\vec k_2$ are the transverse
momenta ($\vec k_1+\vec k_2=\vec q$) of the produced gluon and quark,
respectively.

We need to consider separately the ``inclusive'' situations when either
the quark or the gluon generate the jet, with the kinematics of the other
parton taken arbitrary. We denote the corresponding contributions as
$I^R_{q;q}$ and $I^R_{q;g}$,
\[
I^R_q=I^R_{q;q}+I^R_{q;g} \; .
\]
We start with the case of inclusive jet generation by the gluon, $I_{q;g}^R$.

{\bf a) gluon ``inclusive'' jet generation}

The jet variables are $\vec k=\vec k_1$, $\zeta=\beta_1$
($\beta_2=\bar \zeta\equiv 1-\zeta$, $\vec k_2=\vec q-\vec k$), therefore we
have
\[
I^R_{q;g}=\frac{\alpha_s}{2\pi(4\pi)^\epsilon}
\int\frac{ d^{2+2\epsilon}\vec q}{\pi^{1+\epsilon}}\left(\vec q^{\,\,2}
\right)^{\gamma-{n \over 2}}
\left(\vec q \cdot \vec l \,\, \right)^n \int\limits^1_{\alpha}\frac{d\zeta}
{\zeta}\sum_{a=q,\bar q} f_a\left(\frac{\alpha}{ \zeta}\right)
\]
\beq
\times\frac{1+\bar \zeta^2 +\epsilon \zeta^2}{\zeta}
\left[C_F\frac{1}{\left(\vec q-\frac{\vec k}{\zeta}\right)^2}
+C_A\frac{\bar \zeta}{\zeta}\frac{\frac{\vec k^{\,2}}{\zeta}-\vec k \cdot
\vec q}{(\vec q-\vec k)^2\left(\vec q-\frac{\vec k}{\zeta}\right)^2}\right]\;.
\label{QGgluon}
\eeq

It is worth stressing the difference between the previous calculations of
NLO inclusive parton impact factors and the present case of production of a jet
with fixed momentum. In the parton impact factor case, one keeps
fixed the Reggeon transverse momentum $\vec q$ and integrates over the
allowed phase space of the produced partons, i.e. the integration is
of the form $\int \frac{d\zeta}{2\zeta(1-\zeta)}d^{2+2\epsilon}\vec k\dots$
In the jet production case, instead, we keep fixed the
momentum of the parent parton $\zeta, \vec k$, and allow the Reggeon momentum
$\vec q$ to vary. Indeed, the expression~(\ref{QGgluon}) contains the explicit
integration over the momentum $\vec q$ with the LO BFKL eigenfunctions,
which is needed in order to obtain the impact factor in the
$(\nu,n)$-representation.

The $\vec q$-integration in (\ref{QGgluon}) generates $1/\epsilon$ poles due
to the integrand singularities at $\vec q\to \vec k/\zeta$ for the
contribution proportional to $C_F$ and at $\vec q\to \vec k$ for the one
proportional to $C_A$.
Accordingly we split the result of the $\vec q$-integration into the sum of
two terms: ``singular'' and ``non-singular'' parts. The non-singular part is
defined as
\[
\frac{\alpha_s}{2\pi(4\pi)^\epsilon} \int\limits^1_{\alpha}
\frac{d\zeta}{\zeta}\sum_{a=q,\bar q} f_a\left(\frac{\alpha}{ \zeta}\right)C_A
\frac{\bar \zeta}{\zeta}\left(\frac{1+\bar \zeta^2 +\epsilon \zeta^2}{\zeta}
\right)
\]
\[
\times
\int \frac{d^{2+2\epsilon}\vec q}{\pi^{1+\epsilon}}\frac{\frac{\vec k^{\,2}}
{\zeta}-\vec k \cdot \vec q}{(\vec q-\vec k)^2\left(\vec q-\frac{\vec k}{\zeta}
\right)^2}
\left[\left(\vec q^{\,\,2} \right)^{\gamma-{n \over 2}}
\left(\vec q \cdot \vec l \,\, \right)^n-\left(\vec k^{\,2}
\right)^{\gamma-{n \over 2}}\left(\vec k \cdot \vec l \,\, \right)^n\right]
\]
\[
=\frac{\alpha_s}{2\pi(4\pi)^\epsilon}
 \int\limits^1_{\alpha} \frac{d\zeta}{\zeta}\sum_{a=q,\bar q}
f_a\left(\frac{\alpha}{ \zeta}\right)C_A
\frac{\bar \zeta}{\zeta}\left(\frac{1+\bar \zeta^2 +\epsilon \zeta^2}{\zeta}
\right)
\left(\vec k^{\,2} \right)^{\gamma+\epsilon-{n \over 2}}\left(\vec k \cdot
\vec l \,\, \right)^n
\]
\[
\times\int \frac{d^{2+2\epsilon}\vec a}{\pi^{1+\epsilon}}
\frac{\frac{1}{\zeta}-\vec n \cdot \vec a}{(\vec a-\vec n)^2\left(\vec a
-\frac{\vec n}{\zeta}\right)^2}\left[\left(\vec a^{\,2}
\right)^{\gamma-{n \over 2}}\left(\frac{\vec a \cdot \vec l }
{\vec n \cdot \vec l \,\, }\right)^n-1\right]\;,
\]
where $\vec n$ is a unit vector, $\vec n^{\, 2}=1$. Taking this expression
for $\epsilon=0$ we have
\[
\left(I^R_{q;g}\right)_r=\frac{\alpha_s}{2\pi}\left(\vec k^{\,2}
\right)^{\gamma-{n \over 2}}\left(\vec k \cdot \vec l \,\, \right)^n
 \int\limits^1_{\alpha} \frac{d\zeta}{\zeta}\sum_{a=q,\bar q}
f_a\left(\frac{\alpha}{ \zeta}\right)C_A
\frac{\bar \zeta}{\zeta}\left(\frac{1+\bar \zeta^2 }{\zeta}\right)I_1 \;,
\]
where we define the function
\[
I_1= I_1(n,\gamma,\zeta)=\int \frac{d^{2}\vec a}{\pi}
\frac{\frac{1}{\zeta}-\vec n \cdot \vec a}{(\vec a-\vec n)^2
\left(\vec a-\frac{\vec n}{\zeta}\right)^2}
\left[\left(\vec a^{\,2} \right)^{\gamma}e^{in\phi}-1\right]\;,
\]
with the azimuth $\phi$ of the vector $\vec a$ counted from the
direction of the unit vector $\vec n$.

For the singular contribution we obtain
\[
\frac{\alpha_s}{2\pi}\frac{\Gamma[1-\epsilon]}{(4\pi)^\epsilon}
\frac{1}{\epsilon}\frac{\Gamma^2(1+\epsilon)}{\Gamma(1+2\epsilon)}
\left(\vec k^{\,2} \right)^{\gamma+\epsilon-{n \over 2}}
\left(\vec k \cdot \vec l \,\, \right)^n\int\limits^1_{\alpha}
\frac{d\zeta}{\zeta}\sum_{a=q,\bar q} f_a\left(\frac{\alpha}{ \zeta}\right)
\]
\[
\times\frac{1+\bar \zeta^2+\epsilon\zeta^2}{\zeta}\left[
C_F\frac{\Gamma(1+2\epsilon)\Gamma(\frac{n}{2}-\gamma-\epsilon)
\Gamma(\frac{n}{2}+1+\gamma+\epsilon)}
{\Gamma(1+\epsilon)\Gamma(1-\epsilon)\Gamma(\frac{n}{2}-\gamma)
\Gamma(\frac{n}{2}+1+\gamma+2\epsilon)}
\zeta^{-2\epsilon-2\gamma}+C_A\left(\frac{\bar \zeta}{\zeta}\right)^{2\epsilon}
\right]\;.
\]
Expanding it in $\epsilon$ we get
\[
\left(I^R_{q;g}\right)_s=\frac{\alpha_s}{2\pi}\frac{\Gamma[1-\epsilon]}
{\epsilon (4\pi)^\epsilon}\frac{\Gamma^2(1+\epsilon)}{\Gamma(1+2\epsilon)}
\left(\vec k^{\,\,2} \right)^{\gamma+\epsilon-{n \over 2}}
\left(\vec k \cdot \vec l \,\, \right)^n
\int\limits^1_{\alpha}  \frac{d\zeta}{\zeta} \sum_{a=q,\bar q}f_a
\left(\frac{\alpha}{ \zeta}\right)
\]
\beq
\times \left\{
P_{gq}(\zeta)\left[\frac{C_A}{C_F}+\zeta^{-2\gamma}\right]
\right.
\eeq
\[
\left.
+ \epsilon \left(
\frac{1+\bar \zeta^2}{\zeta}\left[C_F\zeta^{-2\gamma}(\chi(n,\gamma)-2\ln\zeta)
+2C_A\ln\frac{\bar \zeta}{\zeta}\right]+\zeta(C_F\zeta^{-2\gamma}+C_A)
\right)\right\}\;,
\]
where
\beq
\chi(n,\gamma)=2\psi(1)-\psi\left(\frac{n}{2}-\gamma\right)
-\psi\left(\frac{n}{2}+1+\gamma\right)
\label{bfkl-eigenvalue}
\eeq
is the eigenvalue of the LO BFKL kernel,  up to the factor $N\alpha_s/\pi$.

{\bf b) quark ``inclusive'' jet generation}

Now the jet variables are $\vec k=\vec k_2$, $\zeta=\beta_2$ ($\beta_1=\bar
\zeta$, $\vec k_1=\vec q-\vec k$). The corresponding contribution reads
\[
I^R_{q;q}=\frac{\alpha_s}{2\pi(4\pi)^\epsilon}\int \frac{d^{2+2\epsilon}\vec q}
{\pi^{1+\epsilon}}\left(\vec q^{\,\,2} \right)^{\gamma-{n \over 2}}
\left(\vec q \cdot \vec l \,\, \right)^n \int\limits^1_{\alpha} \frac{d\zeta}
{\zeta} \sum_{a=q,\bar q} f_a\left(\frac{\alpha}{ \zeta}\right)
\]
\beq
\times\frac{1+ \zeta^2 +\epsilon \bar \zeta^2}{(1- \zeta)}
\left[
C_F\frac{\bar \zeta^2}{\zeta^2}\frac{\vec k^{\,2}}{(\vec q-\vec k)^2
\left(\vec q-\frac{\vec k}{\zeta}\right)^2}
+C_A\frac{\vec q^{\,\,2}-\vec k \cdot \vec q\, \frac{1+\zeta}{\zeta}
+\frac{\vec k^{\,2}}{\zeta}}{(\vec q-\vec k\,)^2\left(\vec q-\frac{\vec k}
{\zeta}\right)^2}\right]\;.
\label{QGquark}
\eeq

We will consider separately the contributions proportional to $C_F$ and $C_A$.

{\bf b$_1$) quark ``inclusive'' jet generation: $C_F$-term}

Note that the integrand of the $C_F$-term is not singular at $\zeta\to 1$.
We use the decomposition
\[
\frac{\vec k^{\,2}}{(\vec q-\vec k)^2\left(\vec q-\frac{\vec k}{\zeta}
\right)^2} =
\frac{\vec k^{\,2}}{(\vec q-\vec k)^2+\left(\vec q-\frac{\vec k}{\zeta}
\right)^2}\left(\frac{1}{(\vec q-\vec k)^2}
+\frac{1}{\left(\vec q-\frac{\vec k}{\zeta}\right)^2}\right)
\]
in order to separate the regular and singular contributions. The regular part
is given by
\[
\frac{\alpha_s}{2\pi(4\pi)^\epsilon}
\left(\vec k^{\,2} \right)^{\gamma+\epsilon-{n \over 2}}
\left(\vec k \cdot \vec l \,\, \right)^n \int\limits^1_{\alpha}
\frac{d\zeta}{\zeta}\sum_{a=q,\bar q} f_a\left(\frac{\alpha}{ \zeta}\right)
\frac{\bar \zeta}{\zeta}\left(\frac{1+ \zeta^2 +\epsilon \bar \zeta^2}
{\zeta}\right)
\]
\beq
\times
C_F\int \frac{d^{2+2\epsilon}\vec a}{\pi^{1+\epsilon}}
\frac{1}{(\vec a-\vec n)^2+\left(\vec a-\frac{\vec n}{\zeta}\right)^2}
\left[\frac{\left(\vec a^{\,2} \right)^{\gamma-{n \over 2}}
\left(\frac{\vec a \cdot \vec l }{\vec n \cdot \vec l \,\, }\right)^n-1}
{(\vec a-\vec n)^2}+
\frac{\left(\vec a^{\,2} \right)^{\gamma-{n \over 2}}
\left(\frac{\vec a \cdot \vec l }{\vec n \cdot \vec l \,\, }\right)^n
-\zeta^{-2\gamma}}{(\vec a-\frac{\vec n}{\zeta})^2} \right]\;.
\label{regCF}
\eeq
Therefore for $\epsilon=0$ we have
\beq
\left(I_q^q\right)_r^{C_F}=
\frac{\alpha_s}{2\pi}\left(\vec k^{\,2} \right)^{\gamma-{n \over 2}}
\left(\vec k \cdot \vec l \,\, \right)^n
 \int\limits^1_{\alpha}  \frac{d\zeta}{\zeta}\sum_{a=q,\bar q}
f_a\left(\frac{\alpha}{ \zeta}\right)\frac{\bar \zeta (1+ \zeta^2 )}{\zeta^2}
C_F I_2\;,
\eeq
where we define the function
\beq
I_2=I_2(n,\gamma,\zeta)=\int \frac{d^{2}\vec a}{\pi}
\frac{1}{(\vec a-\vec n)^2+\left(\vec a-\frac{\vec n}{\zeta}\right)^2}
\left[\frac{\left(\vec a^{\,2} \right)^{\gamma}e^{i n \phi}-1}
{(\vec a-\vec n)^2}+
\frac{\left(\vec a^{\,2} \right)^{\gamma}e^{i n \phi}-\zeta^{-2\gamma}}
{(\vec a-\frac{\vec n}{\zeta})^2} \right]\;.
\label{regCF-fin}
\eeq

The singular part is proportional to the integral
\[
\int \frac{d^{2+2\epsilon}\vec q}{\pi^{1+\epsilon}} \frac{\vec k^{\,2}}
{(\vec q-\vec k)^2+\left(\vec q-\frac{\vec k}{\zeta}\right)^2}
\left[\frac{\left(\vec k^{\,2} \right)^{\gamma-{n \over 2}}
\left(\vec k \cdot \vec l \,\, \right)^n}{(\vec q-\vec k)^2}
+\frac{\left(\left({\vec k\over \zeta }\right)^2
\right)^{\gamma-{n \over 2}}\left({\vec k \over \zeta} \cdot \vec l \,\,
\right)^n}{(\vec q-{\vec k\over \zeta})^2}\right]
\]
\[
=\frac{\Gamma(1-\epsilon)\Gamma^2(1+\epsilon)}{\epsilon\Gamma[1+2\epsilon]}
\left(\vec k^{\,2} \right)^{\gamma+\epsilon-{n \over 2}}
\left(\vec k \cdot \vec l \,\, \right)^n \left(\frac{\bar \zeta}{\zeta}
\right)^{2\epsilon-2}\left(1+\zeta^{-2\gamma}\right)\;,
\]
therefore, for the singular part of the $C_F$-term we have
\[
\frac{\alpha_s}{2\pi}\frac{\Gamma(1-\epsilon)}{\epsilon (4\pi)^\epsilon}
\frac{\Gamma^2(1+\epsilon)}{\Gamma(1+2\epsilon)}\int\limits^1_{\alpha}
\frac{d\zeta}{\zeta }\left(\vec k^{\,2} \right)^{\gamma+\epsilon-{n \over 2}}
\left(\vec k \cdot \vec l \,\, \right)^n \sum_{a=q,\bar q}
f_a\left(\frac{\alpha}{\zeta }\right)
\]
\[
\times C_F\frac{\left(1+ \zeta^2 +\epsilon \bar \zeta^2\right)}{(1- \zeta)}
\left(\frac{\bar \zeta}{\zeta}\right)^{2\epsilon}\left(1+\zeta^{-2\gamma}
\right) \;.
\]
The next step is to introduce the plus-prescription, which is defined as
\beq
\int\limits^1_a d \zeta \frac{F(\zeta)}{(1-\zeta)_+} =\int\limits^1_a d \zeta
\frac{F(\zeta)-F(1)}{(1-\zeta)}-\int\limits^a_0 d \zeta \frac{F(1)}{(1-\zeta)}
\; ,
\label{plus}
\eeq
for any function $F(\zeta)$, regular at $\zeta=1$. Note that
\[
(1-\zeta)^{2\epsilon-1}=(1-\zeta)^{2\epsilon -1}_+
+\frac{1}{2\epsilon}\delta(1-\zeta)=\frac{1}{2\epsilon}\delta(1-\zeta)
+\frac{1}{(1-\zeta)_+}+2\epsilon\left(\frac{\ln (1-\zeta)}{1-\zeta}\right)_+
+{\cal O}(\epsilon^2) \; .
\]
Using this result, one can write
\[
C_F \frac{\left(1+ \zeta^2 +\epsilon \bar \zeta^2\right)}{(1-\zeta)}
\left(\frac{\bar \zeta}{\zeta}\right)^{2\epsilon}\left(1+\zeta^{-2\gamma}
\right)
= C_F\left[\frac{2}{\epsilon}\delta(1-\zeta)+ \frac{1+\zeta^2}{(1-\zeta)_+}
\left(1+\zeta^{-2\gamma}\right) \right.
\]
\[
\left.
+\epsilon(1+\zeta^{-2\gamma})\left(\bar \zeta+2(1+\zeta^2)
\left(\frac{\ln (1-\zeta)}{1-\zeta}\right)_+ -2(1+ \zeta^2)\frac{\ln \zeta}
{(1-\zeta)}\right)+{\cal O}(\epsilon^2 )\right]
\]
\[
= C_F\left[\left(\frac{2}{\epsilon}-3\right)\delta(1-\zeta)
+\left( \frac{1+\zeta^2}{(1-\zeta)_+}+ \frac{3}{2}\delta(1-\zeta)\right)
\left(1+\zeta^{-2\gamma}\right) +{\cal O}(\epsilon )\right]
\]
\[
= C_F\left(\frac{2}{\epsilon}-3\right)\delta(1-\zeta)+P_{qq}(\zeta)
\left(1+\zeta^{-2\gamma}\right) +{\cal O}(\epsilon)\;.
\]
Taking this into account and expanding in $\epsilon$ the singular part
of the $C_F$-term, one gets the following result for the divergent
contribution:
\bea
&&
\left(I^R_{q;q}\right)_s^{C_F}=\frac{\alpha_s}{2\pi}\frac{\Gamma[1-\epsilon]}
{\epsilon (4\pi)^\epsilon}\frac{\Gamma^2(1+\epsilon)}{\Gamma(1+2\epsilon)}
\left(\vec k^{\,2} \right)^{\gamma+\epsilon-{n \over 2}}
\left(\vec k \cdot \vec l \,\, \right)^n
\int\limits^1_{\alpha}  \frac{d\zeta}{\zeta} \sum_{a=q,\bar q}f_a
\left(\frac{\alpha}{ \zeta}\right) \nonumber \\
&&
\times \left\{C_F\left(\frac{2}{\epsilon}-3\right)\delta(1-\zeta)
+P_{qq}(\zeta)\left(1+\zeta^{-2\gamma}\right)\right.
\nonumber \\
&&
\left.
+ \, \epsilon \, C_F\, (1+\zeta^{-2\gamma})\, \left(\bar \zeta+2(1+\zeta^2)
\left(\frac{\ln (1-\zeta)}{1-\zeta}\right)_+ -2(1+ \zeta^2)
\frac{\ln \zeta}{(1-\zeta)} \right) \right\}\;.
\eea

{\bf b$_2$) quark ``inclusive'' jet generation: $C_A$-term}

The $C_A$-contribution needs a special treatment due to the behavior
of~(\ref{QGquark}) in the region $\zeta\to 1$. 
We use the following decomposition:
\[
\frac{C_A}{(1- \zeta)}
\left(\frac{\vec q^{\,\,2}-\vec k \cdot \vec q\, \frac{1+\zeta}{\zeta}
+\frac{\vec k^{\,2}}{\zeta}}{(\vec q-\vec k\,)^2\left(\vec q-\frac{\vec k}
{\zeta}\right)^2}\right)=\frac{C_A}{2}\frac{2}{(\vec q-\vec k)^2}
\frac{1}{(1- \zeta )}
\]
\[
+\frac{C_A}{2(1-\zeta)}\left[\frac{1}{(\vec q-\frac{\vec k}{\zeta})^2}
-\frac{1}{(\vec q-\vec k)^2}
-\left(\frac{\bar \zeta}{\zeta}\right)^2\frac{\vec k^{\,2}}{(\vec q-\vec k)^2
(\vec q-\frac{\vec k}{\zeta})^2}\right] \;.
\]
The second term in the r.h.s. is regular for $\zeta\to 1$ and can be treated
similarly to what we did above in the case of the $C_F$-contribution.
The first term is singular and the integration over $\zeta$ has to be
restricted, according to definition of NLO impact factor, see
Eq.~(\ref{eq:a19}), by the requirement
\[
M^2_{QG}\leq s_\Lambda\, ,\quad M^2_{QG}=\frac{\vec k_1^{\,2}}{\beta_1}
+\frac{\vec k_2^{\,2}}{\beta_2}-\vec q^{\,\, 2}=
\frac{(\vec q-\vec k)^2}{1-\zeta}+\frac{\vec k^{\,2}}{\zeta}-\vec q^{\,\, 2}\;,
\]
and assuming the $s_\Lambda$ parameter to be much larger than any scale
involved, $s_\Lambda\gg \vec q^{\, 2}, \vec k^{\,\, 2}_{1,2}$. Therefore the
$\zeta$ integral has the form
\[
\int\limits^{1-\zeta_0}_{a} d \zeta \frac{F(\zeta)}{1-\zeta}\, , \quad
{\rm for} \quad \zeta_0=\frac{(\vec q-\vec k)^2}{s_\Lambda}\to 0\; .
\]
Using the plus-prescription (\ref{plus}) one can write
\beq
\int\limits^{1-\zeta_0}_{a} d \zeta \frac{F(\zeta)}{1-\zeta}
=\int\limits^1_a d \zeta \frac{F(\zeta)}{(1-\zeta)_+}
+ F(1)\ln\frac{1}{\zeta_0} \, , \quad {\rm for} \quad \zeta_0\to 0\;,
\label{zeta0}
\eeq
for any function $F(\zeta)$ not singular in the limit $\zeta\to 1$, and
\[
\frac{C_A}{(1- \zeta)}
\left(\frac{\vec q^{\,\,2}-\vec k \cdot \vec q\, \frac{1+\zeta}{\zeta}
+\frac{\vec k^{\,2}}{\zeta}}{(\vec q-\vec k\,)^2\left(\vec q-\frac{\vec k}
{\zeta}\right)^2}\right)=\frac{C_A}{2}\delta(1-\zeta)\frac{2}
{(\vec q-\vec k)^2}\ln\frac{s_\Lambda}{(\vec q-\vec k)^2 }
\]
\[
+\frac{C_A}{2}\frac{2}{(\vec q-\vec k)^2}\frac{1}{(1- \zeta)_+ }+
\frac{C_A}{2(1-\zeta)}\left[\frac{1}{(\vec q-\frac{\vec k}{\zeta})^2}
-\frac{1}{(\vec q-\vec k)^2}-\left(\frac{\bar \zeta}{\zeta}\right)^2
\frac{\vec k^{\,2}}{(\vec q-\vec k)^2(\vec q-\frac{\vec k}{\zeta})^2}\right]\;.
\]
We remind that the definition of NLO impact factor requires the
subtraction of the contribution coming from the gluon emission in the central
rapidity region, given by the last term in Eq.~(\ref{eq:a19}), which we call
below ``BFKL subtraction term''. After this subtraction the parameter
$s_\Lambda$ should be sent to infinity, $s_\Lambda\to \infty$. Our simple
treatment of the invariant mass constraint, $M^2_{QG}\leq s_\Lambda$,
anticipates this limit $s_\Lambda\to \infty$, therefore we neglect all
contributions which are suppressed by powers of $1/s_\Lambda$. Moreover,
the first term in the r.h.s. of the above equation should be naturally
combined with the BFKL subtraction term, giving finally
\[
\frac{C_A}{(1- \zeta)}
\left(\frac{\vec q^{\,\,2}-\vec k \cdot \vec q\, \frac{1+\zeta}{\zeta}
+\frac{\vec k^{\,2}}{\zeta}}{(\vec q-\vec k\,)^2\left(\vec q-\frac{\vec k}
{\zeta}\right)^2}\right)\to\frac{C_A}{2}\delta(1-\zeta)
\frac{1}{(\vec q-\vec k)^2}\ln\frac{s_0}{(\vec q-\vec k)^2 }
\]
\[
+\frac{C_A}{2}\frac{2}{(\vec q-\vec k)^2}\frac{1}{(1- \zeta)_+ }+
\frac{C_A}{2(1-\zeta)}\left[\frac{1}{(\vec q-\frac{\vec k}{\zeta})^2}
-\frac{1}{(\vec q-\vec k)^2}-\left(\frac{\bar \zeta}{\zeta}\right)^2
\frac{\vec k^{\,2}}{(\vec q-\vec k)^2(\vec q-\frac{\vec k}{\zeta})^2}\right]\;,
\]
a result where the artificial parameter $s_\Lambda$ cancels out, as expected.

After that, we are ready to perform the $\vec q$-integration, which naturally
introduces the separation into singular and non-singular contributions. The
singular contribution reads
\[
\frac{\alpha_s}{2\pi}\frac{\Gamma(1-\epsilon)}{\epsilon (4\pi)^\epsilon}
\frac{\Gamma^2(1+\epsilon)}{\Gamma(1+2\epsilon)}
\left(\vec k^{\,2} \right)^{\gamma+\epsilon-{n \over 2}}
\left(\vec k \cdot \vec l \,\, \right)^n\int\limits^1_{\alpha}
\frac{d\zeta}{\zeta } \sum_{a=q,\bar q}f_a\left(\frac{\alpha}{\zeta }\right)
\]
\[
\times  \frac{C_A}{2}\left(1+ \zeta^2 +\epsilon \bar \zeta^2\right)
\left\{\frac{\Gamma(1+2\epsilon)\Gamma(\frac{n}{2}-\gamma-\epsilon)
\Gamma(\frac{n}{2}+1+\gamma+\epsilon)}{\Gamma(1+\epsilon)\Gamma(1-\epsilon)
\Gamma(\frac{n}{2}-\gamma)\Gamma(\frac{n}{2}+1+\gamma+2\epsilon)}\right.
\]
\[
\times\left.\left[
\delta(1-\zeta)\left(\ln\frac{s_0}{\vec k^{\,2}}+\psi\left(\frac{n}{2}-\gamma
-\epsilon\right)
+\psi\left(1+\gamma+\frac{n}{2}+2\epsilon\right)-\psi(\epsilon)-\psi(1)
\right)+\frac{2}{(1-\zeta)_+ }\right.\right.
\]
\[
\left.\left.
+\frac{(\zeta^{-2\epsilon-2\gamma}-1)}{1-\zeta}\right]
-\bar \zeta^{2\epsilon-1}\left(\zeta^{-2\epsilon}+\zeta^{-2\gamma-2\epsilon}
\right)\right\}\;.
\]
Expanding this expression in $\epsilon$ and using that
\[
\frac{(\zeta^{-2\epsilon-2\gamma}-1)}{1-\zeta}=
\frac{(\zeta^{-2\epsilon-2\gamma}-1)}{(1-\zeta)_+}
\]
and
\[
\bar \zeta^{2\epsilon-1}\left(\zeta^{-2\epsilon}+\zeta^{-2\gamma-2\epsilon}
\right)=
\left(\zeta^{-2\epsilon}+\zeta^{-2\gamma-2\epsilon}\right)
\left(\frac{\delta(1-\zeta)}{2\epsilon}+\frac{1}{(1-\zeta)_+}
+{\cal O}(\epsilon)\right)\;,
\]
we get the divergent term
\bea
&&
\left(I^R_{q;q}\right)_s^{C_A}=\frac{\alpha_s}{2\pi}\frac{\Gamma(1-\epsilon)}
{\epsilon (4\pi)^\epsilon}\frac{\Gamma^2(1+\epsilon)}{\Gamma(1+2\epsilon)}
\left(\vec k^{\,2} \right)^{\gamma+\epsilon-{n \over 2}}
\left(\vec k \cdot \vec l \,\, \right)^n \int\limits^1_{\alpha}
\frac{d\zeta}{\zeta } \sum_{a=q,\bar q}f_a\left(\frac{\alpha}{\zeta }\right)
\nonumber \\
&&
\times \left\{C_A\delta(1-\zeta)\ln\frac{s_0}{\vec k^{\,2}} \right. \nonumber \\
&&
+ \, \epsilon \, C_A \left[\delta(1-\zeta)\left(\chi(n,\gamma)
\ln\frac{s_0}{\vec k^{\,2}}+\frac{1}{2}\left(\psi^\prime
\left(1+\gamma+\frac{n}{2}\right)
-\psi^\prime\left(\frac{n}{2}-\gamma\right)-\chi^2(n,\gamma)\right)
\right)\right.\nonumber \\
&&
\left.\left.
+\, (1+\zeta^2)\left((1+\zeta^{-2\gamma})\left(\frac{\chi(n,\gamma)}
{2(1-\zeta)_+}-\left(\frac{\ln (1-\zeta)}{1-\zeta}\right)_+\right)
+\frac{\ln\zeta}{(1-\zeta)} \right)\right]\right\}\;.
\eea

The regular contribution differs from~(\ref{regCF}) only by one
factor and reads
\[
\left(I^R_{q;q}\right)_r^{C_A}=\frac{\alpha_s}{2\pi}\left(\vec k^{\,2}
\right)^{\gamma-{n \over 2}}\left(\vec k \cdot \vec l \,\, \right)^n
\int\limits^1_{\alpha}  \frac{d\zeta}{\zeta}\sum_{a=q,\bar q} f_a
\left(\frac{\alpha}{ \zeta}\right)
\frac{\bar \zeta}{\zeta}\left(\frac{1+ \zeta^2 }{\zeta}\right)
\left(-\frac{C_A}{2}\right) I_2\;.
\]

{\bf c) both quark and gluon generate the jet}

In this case the jet momentum is $\vec k=\vec k_1+\vec k_2$ and the jet
fraction is $1=\zeta+\bar \zeta$. Introducing the vector $\vec \Delta$ as
\[
\vec k_1=\zeta \vec k+\vec \Delta \;,
\]
the contribution reads
\[
I^R_{q;q+g}=
\frac{\alpha_s}{2\pi(4\pi)^\epsilon}\left(\vec k^{\,2}
\right)^{\gamma-{n \over 2}}\left(\vec k \cdot \vec l \,\, \right)^n
\sum_{a=q,\bar q} f_a\left(\alpha\right)\int\frac{ d^{2+2\epsilon}\vec \Delta}
{\pi^{1+\epsilon}}\int\limits^1_{0} d\zeta
\]
\beq
\times\frac{1+\bar \zeta^2 +\epsilon \zeta^2}{\zeta}\left[
C_F\frac{\zeta^2 \vec k^{\,2}}{\vec \Delta^2(\zeta\vec k+\vec \Delta)^2}
+C_A \frac{\bar \zeta \vec k^{\,2}(\zeta \vec k\cdot\vec \Delta+\vec\Delta^2) }
{\vec \Delta^2(\zeta\vec k+\vec \Delta)^2(\bar \zeta\vec k-\vec \Delta)^2}
\right]\;.
\label{example1}
\eeq

In the small-cone approximation (SCA) we need to consider only
\beq
\frac{\alpha_s}{2\pi(4\pi)^\epsilon}\left(\vec k^{\,2}
\right)^{\gamma-{n \over 2}}\left(\vec k \cdot \vec l \,\, \right)^n
\sum_{a=q,\bar q} f_a\left(\alpha\right)C_F \int\limits^1_{0} d\zeta
\frac{1+\bar \zeta^2 +\epsilon \zeta^2}{\zeta}
\int\limits^{\vec\Delta^2_{\rm{max}}}\frac{ d^{2+2\epsilon}\vec \Delta}
{\pi^{1+\epsilon}}\frac{1}{\vec \Delta^2}\;,
\label{example2}
\eeq
where $|\vec \Delta_{\rm{max}}|=|\vec k| R \min (\zeta,\bar \zeta)$. Using that
\[
\int\limits^{\vec \Delta^2_{\rm{max}}}\frac{ d^{2+2\epsilon}\vec \Delta}
{\pi^{1+\epsilon}}\frac{1}{\vec \Delta^2}=\frac{1}{\epsilon\Gamma(1+\epsilon)}
(\vec\Delta_{\rm{max}}^2)^{\epsilon}\approx\frac{\Gamma(1-\epsilon)
\Gamma^2(1+\epsilon)}{\epsilon\Gamma(1+2\epsilon)}
(\vec\Delta_{\rm{max}}^2)^{\epsilon}\;,
\]
we get
\[
I^R_{q;q+g}=
\frac{\alpha_s}{2\pi}\frac{\Gamma(1-\epsilon)}{\epsilon (4\pi)^\epsilon}
\frac{\Gamma^2(1+\epsilon)}{\Gamma(1+2\epsilon)}
\left(\vec k^{\,2} \right)^{\gamma+\epsilon-{n \over 2}}
\left(\vec k \cdot \vec l \,\, \right)^n\sum_{a=q,\bar q} f_a
\left(\alpha\right)C_F R^{2\epsilon}
\]
\beq
\times\int\limits^1_{0}   d\zeta (\min (\zeta,\bar \zeta))^{2\epsilon}
\frac{1+\bar \zeta^2 +\epsilon \zeta^2}{\zeta}
\eeq
\[
\approx
\frac{\alpha_s}{2\pi}\frac{\Gamma(1-\epsilon)}{\epsilon (4\pi)^\epsilon}
\frac{\Gamma^2(1+\epsilon)}{\Gamma(1+2\epsilon)}
\left(\vec k^{\,2} \right)^{\gamma+\epsilon-{n \over 2}}
\left(\vec k \cdot \vec l \,\, \right)^n
\]
\[
\times \sum_{a=q,\bar q} f_a
\left(\alpha\right)C_F R^{2\epsilon}\left[\frac{1}{\epsilon}-\frac{3}{2}
+\epsilon\left(\frac{7}{2}-\frac{\pi^2}{3}+3\ln 2\right)\right]\;.
\]

{\bf d) gluon ``inclusive'' jet generation with the quark in the jet cone}

We introduce the vector $\vec \Delta$ such that
\[
\vec q=\frac{\vec k}{\zeta}+\vec \Delta\;,
\]
where $\vec k$ coincides with $\vec k_1$, the transverse momentum of the gluon
generating the jet. The contribution reads
\[
I^R_{q;g,-q}=
-\frac{\alpha_s}{2\pi(4\pi)^\epsilon}\left(\vec k^{\,2}
\right)^{\gamma-{n \over 2}}\left(\vec k \cdot \vec l \,\, \right)^n
\int\limits^1_{\alpha}   \frac{d\zeta}{\zeta} \zeta^{-2\gamma}
\sum_{a=q,\bar q} f_a\left(\frac{\alpha}{ \zeta}\right)C_F
\]
\beq
\times\frac{1+\bar \zeta^2 +\epsilon \zeta^2}{\zeta}
\int\limits^{\vec \Delta^2_{\rm{max}}}\frac{ d^{2+2\epsilon}\vec \Delta}
{\pi^{1+\epsilon}}\frac{1}{\vec \Delta^2}\;,
\eeq
where now $|\vec\Delta_{\rm{max}}|=|\vec k| R \frac{\bar \zeta}{\zeta}$. We get
\[
I^R_{q;g,-q}=
-\frac{\alpha_s}{2\pi}\frac{\Gamma(1-\epsilon)}{\epsilon (4\pi)^\epsilon}
\frac{\Gamma^2(1+\epsilon)}{\Gamma(1+2\epsilon)}
\left(\vec k^{\,2} \right)^{\gamma+\epsilon-{n \over 2}}
\left(\vec k \cdot \vec l \,\, \right)^nC_F R^{2\epsilon}
\]
\beq
\times\int\limits^1_{\alpha}   \frac{d\zeta}{\zeta} \zeta^{-2\gamma}
\sum_{a=q,\bar q} f_a\left(\frac{\alpha}{\zeta}\right)
\left(\frac{\bar \zeta}{\zeta}\right)^{2\epsilon}
\frac{1+\bar \zeta^2 +\epsilon \zeta^2}{\zeta}
\eeq
\[
\approx
-\frac{\alpha_s}{2\pi}\frac{\Gamma(1-\epsilon)}{\epsilon (4\pi)^\epsilon}
\frac{\Gamma^2(1+\epsilon)}{\Gamma(1+2\epsilon)}
\left(\vec k^{\,2} \right)^{\gamma+\epsilon-{n \over 2}}
\left(\vec k \cdot \vec l \,\, \right)^n R^{2\epsilon}\int\limits^1_{\alpha}
\frac{d\zeta}{\zeta} \zeta^{-2\gamma} \sum_{a=q,\bar q} f_a
\left(\frac{\alpha}{\zeta}\right)
\]
\[
\times\left( P_{gq}(\zeta)\left[1+2\epsilon\ln\frac{\bar \zeta}{\zeta}\right]
+\epsilon C_F \zeta \right)\;.
\]
Note the overall minus sign, which means that this contribution is a
subtractive term to the gluon ``inclusive'' jet generation.

{\bf e) quark ``inclusive'' jet generation with the gluon in the jet cone}

In this case we have
\[
\vec q=\frac{\vec k}{\zeta}+\vec \Delta\;,
\]
with $\vec k$ identified with $\vec k_2$, the transverse momentum of the quark
generating the jet. The contribution reads
\[
I^R_{q;q,-g}=
-\frac{\alpha_s}{2\pi(4\pi)^\epsilon}\left(\vec k^{\,2}
\right)^{\gamma-{n \over 2}}\left(\vec k \cdot \vec l \,\, \right)^n
\int\limits^1_{\alpha} \frac{d\zeta}{\zeta}\zeta^{-2\gamma}
\sum_{a=q,\bar q} f_a\left(\frac{\alpha}{ \zeta}\right)C_F
\]
\beq
\times\frac{1+ \zeta^2 +\epsilon \bar \zeta^2}{(1- \zeta)}
\int\limits^{\vec \Delta^2_{\rm{max}}}\frac{ d^{2+2\epsilon}\vec \Delta}
{\pi^{1+\epsilon}}\frac{1}{\vec \Delta^2}\;,
\eeq
where $|\vec\Delta_{\rm{max}}|=|\vec k| R \frac{\bar \zeta}{\zeta}$. We get
\[
I^R_{q;q,-q}=
-\frac{\alpha_s}{2\pi}\frac{\Gamma(1-\epsilon)}{\epsilon (4\pi)^\epsilon}
\frac{\Gamma^2(1+\epsilon)}{\Gamma(1+2\epsilon)}
\left(\vec k^{\,2} \right)^{\gamma+\epsilon-{n \over 2}}
\left(\vec k \cdot \vec l \,\, \right)^nC_F R^{2\epsilon}
\]
\beq
\times\int\limits^1_{\alpha}   \frac{d\zeta}{\zeta} \zeta^{-2\gamma}
\sum_{a=q,\bar q} f_a\left(\frac{\alpha}{\zeta}\right)
\left(\frac{\bar \zeta}{\zeta}\right)^{2\epsilon}
\frac{1+ \zeta^2 +\epsilon \bar \zeta^2}{(1-\zeta)}
\eeq
\[
\approx
-\frac{\alpha_s}{2\pi}\frac{\Gamma(1-\epsilon)}{\epsilon (4\pi)^\epsilon}
\frac{\Gamma^2(1+\epsilon)}{\Gamma(1+2\epsilon)}
\left(\vec k^{\,2} \right)^{\gamma+\epsilon-{n \over 2}}
\left(\vec k \cdot \vec l \,\, \right)^n R^{2\epsilon}\int\limits^1_{\alpha}
   \frac{d\zeta}{\zeta} \zeta^{-2\gamma} \sum_{a=q,\bar q} f_a
\left(\frac{\alpha}{\zeta}\right)
\]
\[
\times \left( P_{qq}(\zeta)+C_F\delta(1-\zeta)\left(\frac{1}{\epsilon}
-\frac{3}{2}\right)+\epsilon C_F \left(\bar \zeta -2\frac{(1+\zeta^2)\ln\zeta}
{(1-\zeta)}+2(1+\zeta^2)\left(\frac{\ln(1-\zeta)}{(1-\zeta)}\right)_+\right)
\right)\;.
\]

\subsubsection{Final result for the case of incoming quark}

Collecting all the contributions calculated in this Section and taking
into account the PDFs' renormalization counterterm~(\ref{c.count.t})
and the charge counterterm~(\ref{charge.count.t}), we find that all singular
contributions cancel and the result is
\beq
I_q=
\frac{\alpha_s}{2\pi}\left(\vec k^{\,2} \right)^{\gamma-{n \over 2}}
\left(\vec k \cdot \vec l \,\, \right)^n
\int\limits^1_{\alpha}  \frac{d\zeta}{\zeta}\sum_{a=q,\bar q} f_a
\left(\frac{\alpha}{ \zeta}\right)
\label{resultq}
\eeq
\[
\times\left[\left\{P_{qq}(\zeta)+\frac{C_A}{C_F}P_{gq}(\zeta)\right\}
\ln\frac{\vec k^{\,2}}{\mu_F^2}-2\zeta^{-2\gamma}\ln R\,
\left\{P_{qq}(\zeta)+P_{gq}(\zeta)\right\}-\frac{\beta_0}{2}
\ln\frac{\vec k^{\,2}}{\mu_R^2}\delta(1-\zeta)\right.
\]
\[
+C_A\delta(1-\zeta)\left\{\chi(n,\gamma)\ln\frac{s_0}{\vec k^{\,2}}
+\frac{85}{18}+\frac{\pi^2}{2}+\frac{1}{2}\left(\psi^\prime
\left(1+\gamma+\frac{n}{2}\right)
-\psi^\prime\left(\frac{n}{2}-\gamma\right)-\chi^2(n,\gamma)\right)
\right\}
\]
\[
+(1+\zeta^2)\left\{C_A\left(\frac{(1+\zeta^{-2\gamma})\,\chi(n,\gamma)}
{2(1-\zeta)_+}-\zeta^{-2\gamma}\left(\frac{\ln(1-\zeta)}{1-\zeta}\right)_+
\right)+\left(C_F-\frac{C_A}{2}\right)\left[ \frac{\bar \zeta}{\zeta^2}I_2
-\frac{2\ln\zeta}{1-\zeta}\right.\right.
\]
\[
\left.\left.
+2\left(\frac{\ln(1-\zeta)}{1-\zeta}\right)_+ \right]\right\}+
\delta(1-\zeta)\left(C_F\left(3\ln 2-\frac{\pi^2}{3}-\frac{9}{2}\right)
-\frac{5n_f}{9}\right)
\]
\[
\left. +C_A\zeta+C_F\bar \zeta+\frac{1+\bar \zeta^2}{\zeta}
\left\{C_A\frac{\bar \zeta}{\zeta}I_1+2C_A\ln\frac{\bar\zeta}{\zeta}
+C_F\zeta^{-2\gamma}(\chi(n,\gamma)-2\ln \bar \zeta)\right\}\right]\;.
\]

\subsection{Incoming gluon}

We distinguish virtual corrections and real emission contributions,
\[
I_g=I_g^V+I_g^R \; .
\]

Virtual corrections are the same as in the case of inclusive gluon impact
factor,
\[
I^V_g=-\frac{\alpha_s}{2\pi}\frac{\Gamma[1-\epsilon]}{(4\pi)^\epsilon}
\frac{1}{\epsilon}\frac{\Gamma^2(1+\epsilon)}{\Gamma(1+2\epsilon)}
\left(\vec k^{\,2} \right)^{\gamma+\epsilon-{n \over 2}}
\left(\vec k \cdot \vec l \,\, \right)^n   f_g(\alpha ) \frac{C_A}{C_F}
\]
\[
\times \left\{C_A\left(\ln\frac{s_0}{\vec k^{\,2}}+\frac{2}{\epsilon}
-\frac{11+9\epsilon}{2(1+2\epsilon)(3+2\epsilon)}+\psi(1-\epsilon)
-2\psi(1+\epsilon)+\psi(1)\right.\right.
\]
\beq
\left.\left.
+\frac{\epsilon}{(1+\epsilon)(1+2\epsilon)(3+2\epsilon)}\right)
+n_f\left(\frac{(1+\epsilon)(2+\epsilon)-1-\frac{\epsilon}{1+\epsilon}}
{(1+\epsilon)(1+2\epsilon)(3+2\epsilon)}\right)\right\} \;.
\label{Gvirt}
\eeq

Expanding it in $\epsilon$ we obtain the following results for the singular,
\[
\left(I^V_g\right)_s=-\frac{\alpha_s}{2\pi}\frac{\Gamma[1-\epsilon]}
{(4\pi)^\epsilon}\frac{1}{\epsilon}\frac{\Gamma^2(1+\epsilon)}
{\Gamma(1+2\epsilon)} \left(\vec k^{\,2} \right)^{\gamma+\epsilon
-{n \over 2}}\left(\vec k \cdot \vec l \,\, \right)^n f_g
(\alpha)\frac{C_A}{C_F}
\]
\beq
\times\left\{C_A\left(\ln\frac{s_0}{\vec k^{\,2}}+\frac{2}{\epsilon}
-\frac{11}{6}\right)+\frac{n_f}{3}\right\}\;,
\label{Gvirt-sing}
\eeq
and the finite parts,
\beq
\left(I^V_g\right)_r=-\frac{\alpha_s}{2\pi}
\left(\vec k^{\,2} \right)^{\gamma-{n \over 2}}
\left(\vec k \cdot \vec l \,\, \right)^n  f_g(\alpha) \frac{C_A}{C_F}
\left\{C_A\left(\frac{67}{18}-\frac{\pi^2}{2}\right)-\frac{5}{9} n_f
\right\} \;.
\label{Gvirt-nsing}
\eeq

For the corrections due to real emissions, one has to consider quark-antiquark
and two-gluon intermediate states,
\[
I_g^R=I^R_{g;q}+I^R_{g;g} \ .
\]

\subsubsection{Quark-antiquark intermediate state}

The starting point here is the quark-antiquark intermediate state contribution
to the inclusive gluon impact factor ($T_R=1/2$),
\beq
\Phi^{\{Q\bar Q\}}=\Phi_g g^2\vec q^{\,\, 2}\frac{d^{2+2\epsilon} \vec k_1}
{(2\pi)^{3+2\epsilon}}d\beta_1 T_R\left(1-\frac{2\beta_1\beta_2}{1+\epsilon}
\right)\left\{\frac{C_F}{C_A}\frac{1}{\vec k_1^{\,2} \vec k_2^{\,2}}
+\beta_1\beta_2\frac{\vec k_1\cdot\vec k_2}{\vec k_1^{\,2} \vec k_2^{\,2}
(\vec k_2\beta_1-\vec k_1 \beta_2)^2} \right\}\;,
\eeq
where $\beta_1$ and $\beta_2$ are the relative longitudinal momenta
($\beta_1+\beta_2=1$) and $\vec k_1$ and $\vec k_2$ are the transverse momenta
($\vec k_1+\vec k_2=\vec q\,\,$) of the produced quark and antiquark,
respectively.

{\bf a) quark ``inclusive'' jet generation}

Taking into account the factor $n_f$ arising from the summation over all active
quark flavors, we have the following contribution:
\[
I^R_{g;q}=\frac{\alpha_s}{2\pi(4\pi)^\epsilon}\int \frac{d^{2+2\epsilon}\vec q}
{\pi^{1+\epsilon}}\left(\vec q^{\,\,2} \right)^{\gamma-{n \over 2}}
\left(\vec q \cdot \vec l \,\, \right)^n  n_f \int\limits^1_{\alpha}
\frac{d\zeta}{\zeta} f_g \left(\frac{\alpha}{ \zeta}\right) \frac{C_A}{C_F}
\]
\beq
\times T_R\left(1-\frac{2\zeta\bar \zeta}{1+\epsilon}\right)
\left\{\frac{C_F}{C_A}\frac{1}{ (\vec q-\vec k)^{ 2}}
+\frac{\bar\zeta}{\zeta} \frac{\vec k\cdot(\vec q-\vec k)}
{(\vec q-\vec k)^{2} (\vec q-\frac{\vec k}{ \zeta})^2} \right\}\;.
\label{QQquark}
\eeq

We can split this integral into the sum of singular and non-singular parts.
For the singular contribution we have
\[
\frac{\alpha_s}{2\pi}\frac{\Gamma[1-\epsilon]}{(4\pi)^\epsilon}
\frac{1}{\epsilon}\frac{\Gamma^2(1+\epsilon)}{\Gamma(1+2\epsilon)}
\left(\vec k^{\,2} \right)^{\gamma+\epsilon-{n \over 2}}
\left(\vec k \cdot \vec l \,\, \right)^n n_f \int\limits^1_{\alpha}
\frac{d\zeta}{\zeta}  f_g \left(\frac{\alpha}{ \zeta}\right)\frac{C_A}{C_F}
\]
\[
\times T_R\left(1-\frac{2\zeta\bar \zeta}{1+\epsilon}\right)
\left[
\frac{C_F}{C_A}
\frac{\Gamma(1+2\epsilon)\Gamma(\frac{n}{2}-\gamma-\epsilon)
\Gamma(\frac{n}{2}+1+\gamma+\epsilon)}
{\Gamma(1+\epsilon)\Gamma(1-\epsilon)\Gamma(\frac{n}{2}-\gamma)
\Gamma(\frac{n}{2}+1+\gamma+2\epsilon)}
+\bar\zeta^{2\epsilon}\zeta^{-2\epsilon-2\gamma}\right]\;.
\]
Expanding it in $\epsilon$ we obtain
\[
\left(I^R_{g;q}\right)_s=\frac{\alpha_s}{2\pi}\frac{\Gamma[1-\epsilon]}
{\epsilon (4\pi)^\epsilon}
\left(\vec k^{\,2} \right)^{\gamma+\epsilon-{n \over 2}}
\left(\vec k \cdot \vec l \,\, \right)^n
\int\limits^1_{\alpha}  \frac{d\zeta}{\zeta} f_g\left(\frac{\alpha}{ \zeta}
\right)\frac{C_A}{C_F} \, n_f
\left\{P_{qg}(\zeta)\left[\frac{C_F}{C_A}+\zeta^{-2\gamma}\right]
\right.
\]
\[
\left.
+\, \epsilon \, \left(2 \zeta \bar \zeta\, T_R
\left[\frac{C_F}{C_A}+\zeta^{-2\gamma}\right]+P_{qg}(\zeta)
\left[\frac{C_F}{C_A}\chi(\gamma,n)+2 \zeta^{-2\gamma}
\ln \frac{\bar \zeta}{\zeta}\right]\right) \right\}\;.
\]

For the regular part of~(\ref{QQquark}) we have
\[
\frac{\alpha_s}{2\pi(4\pi)^\epsilon}\left(\vec k^{\,2}
\right)^{\gamma+\epsilon-{n \over 2}}\left(\vec k \cdot \vec l \,\, \right)^n
n_f\int\limits^1_{\alpha}    \frac{d\zeta}{\zeta}
f_g\left(\frac{\alpha}{ \zeta}\right)\frac{C_A}{C_F}
\]
\[
\times T_R\left(1-\frac{2\zeta\bar \zeta}{1+\epsilon}\right)\frac{\bar \zeta}
{\zeta}\int \frac{d^{2+2\epsilon}\vec a}{\pi^{1+\epsilon}}
\left[\left(\vec a^{\, 2} \right)^{\gamma-\frac{n}{2}}
\left(\frac{\vec a \cdot \vec l \,\,}{\vec n \cdot \vec l \,\,} \right)^n
-\zeta^{-2\gamma}\right]\frac{\vec a\cdot \vec n -1}{(\vec a-\vec n)^2
\left(\vec a-\frac{\vec n}{\zeta}\right)^2}\;.
\]
Expanding it in $\epsilon$ we get
\[
\left(I^R_{g;q}\right)_r=\frac{\alpha_s}{2\pi}\left(\vec k^{\,2}
\right)^{\gamma-{n \over 2}}\left(\vec k \cdot \vec l \,\, \right)^n n_f
\int\limits^1_{\alpha}    \frac{d\zeta}{\zeta}
f_g\left(\frac{\alpha}{ \zeta}\right)\frac{C_A}{C_F}\frac{\bar \zeta}{\zeta}
P_{qg}(\zeta)I_3\;,
\]
where we define the function
\beq
I_3=I_3(n,\gamma,\zeta) =\int \frac{d^{2}\vec a}{\pi}
\frac{\vec a\cdot \vec n -1}{(\vec a-\vec n)^2
\left(\vec a-\frac{\vec n}{\zeta}\right)^2}\left[\left(\vec a^{\, 2}
\right)^{\gamma}e^{in\phi}-\zeta^{-2\gamma}\right]\;.
\eeq

The case of antiquark inclusive generation of the jet is identical to
the case of the quark.

{\bf b) both quark and antiquark generate the jet}

The jet momentum is $\vec k=\vec k_1+\vec k_2$ and the jet fraction is
$1=\zeta+\bar \zeta$. Introducing $\vec \Delta$ as
\[
\vec k_1=\zeta \vec k+\vec \Delta \;,
\]
the contribution reads
\[
I^R_{g;q+\bar q}
=\frac{\alpha_s}{2\pi(4\pi)^\epsilon}\left(\vec k^{\,2} \right)^{\gamma
-{n \over 2}}\left(\vec k \cdot \vec l \,\, \right)^n n_f f_{g}
\left(\alpha\right) \frac{C_A}{C_F}
\int\frac{ d^{2+2\epsilon}\vec \Delta}{\pi^{1+\epsilon}}
\int\limits^1_{0}   d\zeta \, T_R\left(1-\frac{2\zeta\bar \zeta}{1+\epsilon}
\right)
\]
\beq
\times
\left\{\frac{C_F}{C_A}\frac{\vec k^{\,2}}{(\zeta \vec k+\vec \Delta)^2
(\bar \zeta \vec k-\vec \Delta)^2}+\zeta\bar \zeta\ \frac{\vec k^{\,2}
\left(\zeta \vec k+\vec \Delta\right)\cdot\left(\bar \zeta \vec k
-\vec \Delta\right)}{ (\zeta \vec k+\vec \Delta)^2 (\bar \zeta \vec k
-\vec \Delta)^2 \vec \Delta^2 } \right\}\;.
\eeq
In the SCA we need to consider only
\[
\frac{\alpha_s}{2\pi(4\pi)^\epsilon}\left(\vec k^{\,2} \right)^{\gamma
-{n \over 2}}\left(\vec k \cdot \vec l \,\, \right)^n n_f f_{g}
\left(\alpha\right) \frac{C_A}{C_F}
\!\int\limits^1_{0}   d\zeta \, T_R\left(1-\frac{2\zeta\bar \zeta}
{1+\epsilon}\right)
\!\int\limits^{\vec \Delta^2_{\rm{max}}}\frac{ d^{2+2\epsilon}\vec \Delta}
{\pi^{1+\epsilon}}\frac{1}{\vec \Delta^2}\;,
\]
where $|\vec\Delta_{\rm{max}}|=|\vec k| R\min (\zeta,\bar \zeta)$. Using again
that
\[
\int\limits^{\vec \Delta^2_{\rm{max}}}\frac{ d^{2+2\epsilon}\vec \Delta}
{\pi^{1+\epsilon}}\frac{1}{\vec \Delta^2}=\frac{1}{\epsilon\Gamma(1+\epsilon)}
(\vec \Delta^2_{\rm{max}})^\epsilon\approx
\frac{\Gamma(1-\epsilon)\Gamma^2(1+\epsilon)}{\epsilon\Gamma(1+2\epsilon)}
(\vec \Delta^2_{\rm{max}})^\epsilon\;,
\]
we get
\[
I^R_{g;q+\bar q}=
\frac{\alpha_s}{2\pi}\frac{\Gamma(1-\epsilon)}{\epsilon (4\pi)^\epsilon}
\frac{\Gamma^2(1+\epsilon)}{\Gamma(1+2\epsilon)}
\left(\vec k^{\,2} \right)^{\gamma+\epsilon-{n \over 2}}
\left(\vec k \cdot \vec l \,\, \right)^n n_f f_{g}\left(\alpha\right)
\frac{C_A}{C_F} R^{2\epsilon}
\]
\beq
\times\int\limits^1_{0} d\zeta \, T_R(\min (\zeta,\bar \zeta))^{2\epsilon}
\left(1-\frac{2\zeta\bar \zeta}{1+\epsilon}\right)
\eeq
\[
\approx
\frac{\alpha_s}{2\pi}\frac{\Gamma(1-\epsilon)}{\epsilon (4\pi)^\epsilon}
\frac{\Gamma^2(1+\epsilon)}{\Gamma(1+2\epsilon)}
\left(\vec k^{\,2} \right)^{\gamma+\epsilon-{n \over 2}}
\left(\vec k \cdot \vec l \,\, \right)^n \, n_f f_{g}\left(\alpha\right)
\frac{C_A}{C_F} R^{2\epsilon}\left[\frac{1}{3}
-\epsilon \left(\frac{23}{36}+\frac{2}{3}\ln 2\right)\right]\;.
\]

{\bf c) quark ``inclusive'' jet generation with the antiquark in the jet cone}

We introduce the vector $\vec \Delta$ such that
\[
\vec q=\frac{\vec k}{\zeta}+\vec \Delta\;,
\]
where $\vec k$ coincides with $\vec k_1$, the transverse momentum of the quark
generating the jet. The contribution reads
\[
I^R_{q;q,-\bar q}=
-\frac{\alpha_s}{2\pi(4\pi)^\epsilon}\left(\vec k^{\,2}
\right)^{\gamma-{n \over 2}}\left(\vec k \cdot \vec l \,\, \right)^n
\int\limits^1_{\alpha}   \frac{d\zeta}{\zeta} \zeta^{-2\gamma}\, n_f f_{g}
\left(\frac{\alpha}{ \zeta}\right) \frac{C_A}{C_F}
\]
\beq
\times T_R\left(1-\frac{2\zeta\bar \zeta}{1+\epsilon}\right)
\int\limits^{\vec \Delta^2_{\rm{max}}}\frac{ d^{2+2\epsilon}\vec \Delta}
{\pi^{1+\epsilon}}\frac{1}{\vec \Delta^2}\;,
\eeq
where now $|\vec\Delta_{\rm{max}}|=|\vec k| R \frac{\bar \zeta}{\zeta}$. We get
\[
I^R_{q;q,-\bar q}=
-\frac{\alpha_s}{2\pi}\frac{\Gamma(1-\epsilon)}{\epsilon (4\pi)^\epsilon}
\frac{\Gamma^2(1+\epsilon)}{\Gamma(1+2\epsilon)}
\left(\vec k^{\,2} \right)^{\gamma+\epsilon-{n \over 2}}
\left(\vec k \cdot \vec l \,\, \right)^n R^{2\epsilon}
\]
\beq
\times\int\limits^1_{\alpha}   \frac{d\zeta}{\zeta}
\,  f_{g}\left(\frac{\alpha}{ \zeta}\right) \frac{C_A}{C_F} \zeta^{-2\gamma}
\left(\frac{\bar \zeta}{\zeta}\right)^{2\epsilon}
T_R\, n_f\left(1-\frac{2\zeta\bar \zeta}{1+\epsilon}\right)
\eeq
\[
\approx
-\frac{\alpha_s}{2\pi}\frac{\Gamma(1-\epsilon)}{\epsilon (4\pi)^\epsilon}
\frac{\Gamma^2(1+\epsilon)}{\Gamma(1+2\epsilon)}
\left(\vec k^{\,2} \right)^{\gamma+\epsilon-{n \over 2}}
\left(\vec k \cdot \vec l \,\, \right)^n R^{2\epsilon}\int\limits^1_{\alpha}
\frac{d\zeta}{\zeta} f_{g}\left(\frac{\alpha}{ \zeta}\right) \frac{C_A}{C_F}
\]
\[
\times \zeta^{-2\gamma} \, n_f \left( P_{qg}(\zeta)
\left[1+2\epsilon\ln\frac{\bar \zeta}{\zeta}\right]+\epsilon \,
\zeta \bar \zeta \right)\;.
\]
Note the overall minus sign, which means that this contribution is a
subtractive term to the quark ``inclusive'' jet generation.

The case of antiquark ``inclusive'' jet generation with the quark in the jet
cone gives the same contribution.

\subsubsection{Two-gluon intermediate state }

The starting point here is the gluon-gluon intermediate state contribution
to the inclusive gluon impact factor,
\[
\Phi^{\{GG\}}=\Phi_g g^2\vec q^{\,\, 2}\frac{d^{2+2\epsilon} \vec k_1}
{(2\pi)^{3+2\epsilon}}d\beta_1 \frac{C_A}{2}\left[\frac{1}{\beta_1}
+\frac{1}{\beta_2}-2+\beta_1\beta_2\right]
\]
\[
\times\left\{\frac{1}{\vec k_1^{\,2} \vec k_2^{\,2}}
+\frac{\beta_1^2}{\vec k_1^{\,2} (\vec k_2\beta_1-\vec k_1 \beta_2)^2}
+\frac{\beta_2^2}{\vec k_2^{\,2} (\vec k_2\beta_1-\vec k_1 \beta_2)^2}\right\}
\;,
\]
where $\beta_1$ and $\beta_2$ are the relative longitudinal momenta
($\beta_1+\beta_2=1$) and $\vec k_1$ and $\vec k_2$ are the transverse
momenta ($\vec k_1+\vec k_2=\vec q$) of the two produced gluons.

{\bf a) gluon ``inclusive'' jet generation}

We need to consider the case of a gluon which generates the jet, while the
other is a spectator, the case when the other gluon generates the jet
being taken into account by a factor 2. Thus, we obtain the following
integral:
\[
I^R_{g;g}=\frac{\alpha_s}{2\pi(4\pi)^\epsilon}\int d^{2+2\epsilon}\vec q
\left(\vec q^{\,\,2} \right)^{\gamma-{n \over 2}}
\left(\vec q \cdot \vec l \,\, \right)^n  \int\limits^1_{\alpha}
\frac{d\zeta}{\zeta}f_g\left(\frac{\alpha}{ \zeta}\right)\frac{C_A}{C_F}
\]
\beq
\times C_A\left[\frac{1}{\zeta}+\frac{1}{(1- \zeta)}-2+\zeta\bar\zeta\right]
\left\{\frac{1}{ (\vec q-\vec k)^2}+\frac{1}{ \left(\vec q -\frac{\vec k}
{ \zeta}\right)^2} +\frac{\bar \zeta^2}{\zeta^2}\frac{ \vec k^{\,2}}
{(\vec q-\vec k)^2 \left(\vec q -\frac{\vec k}{ \zeta}\right)^2}\right\}\;.
\label{GGgluon}
\eeq

The calculation goes along the same lines as in the Section~\ref{subsub_QG}
(case {\bf b$_2$}). First, we separate the $\zeta\to 1$ singularity,
then we add the BFKL subtraction term. Using~(\ref{zeta0}) one obtains
\[
\left[\frac{1}{\zeta}+\frac{1}{(1- \zeta)}-2+\zeta\bar\zeta\right]
\left\{\frac{1}{ (\vec q-\vec k)^2}+\frac{1}{ \left(\vec q
-\frac{\vec k}{ \zeta}\right)^2} +\frac{\bar \zeta^2}{\zeta^2}
\frac{ \vec k^{\,2}}{(\vec q-\vec k)^2 \left(\vec q
-\frac{\vec k}{ \zeta}\right)^2}\right\}
\]
\[
=\left[\frac{1}{\zeta}-2+\zeta\bar\zeta\right]
\left\{\frac{1}{ (\vec q-\vec k)^2}+\frac{1}{ \left(\vec q
-\frac{\vec k}{ \zeta}\right)^2} +\frac{\bar \zeta^2}{\zeta^2}
\frac{ \vec k^{\,2}}{(\vec q-\vec k)^2 \left(\vec q
-\frac{\vec k}{ \zeta}\right)^2}\right\}
\]
\[
+\frac{1}{(1- \zeta)}\left\{ \frac{1}{ \left(\vec q -\frac{\vec k}{ \zeta}
\right)^2}-\frac{1}{ (\vec q-\vec k)^2}+\frac{\bar \zeta^2}{\zeta^2}
\frac{ \vec k^{\,2}}{(\vec q-\vec k)^2 \left(\vec q
-\frac{\vec k}{ \zeta}\right)^2}\right\}+\frac{1}{(1- \zeta)}
\frac{2}{ (\vec q-\vec k)^2}
\]
\[
\to\left[\frac{1}{\zeta}-2+\zeta\bar\zeta\right]
\left\{\frac{1}{ (\vec q-\vec k)^2}+\frac{1}{ \left(\vec q
-\frac{\vec k}{ \zeta}\right)^2} +\frac{\bar \zeta^2}{\zeta^2}
\frac{ \vec k^{\,2}}{(\vec q-\vec k)^2 \left(\vec q
-\frac{\vec k}{ \zeta}\right)^2}\right\}
\]
\[
+\frac{1}{(1- \zeta)}\left\{ \frac{1}{ \left(\vec q
-\frac{\vec k}{ \zeta}\right)^2}-\frac{1}{ (\vec q-\vec k)^2}
+\frac{\bar \zeta^2}{\zeta^2}\frac{ \vec k^{\,2}}{(\vec q-\vec k)^2
\left(\vec q -\frac{\vec k}{ \zeta}\right)^2}\right\}+\frac{1}{(1- \zeta)_+}
\frac{2}{ (\vec q-\vec k)^2}
\]
\[
+\delta(1-\zeta)\frac{1}{ (\vec q-\vec k)^2}\ln\frac{s_0}{(\vec q-\vec k)^2}\;.
\]

We can split the result into the sum of  singular and non-singular
parts. For the singular contribution we obtain
\[
\frac{\alpha_s}{2\pi}\frac{\Gamma[1-\epsilon]}{(4\pi)^\epsilon}
\frac{1}{\epsilon}\frac{\Gamma^2(1+\epsilon)}{\Gamma(1+2\epsilon)}
\left(\vec k^{\,2} \right)^{\gamma+\epsilon-{n \over 2}}
\left(\vec k \cdot \vec l \,\, \right)^n \int\limits^1_{\alpha}
\frac{d\zeta}{\zeta}f_g\left(\frac{\alpha}{ \zeta}\right)\frac{C_A}{C_F} C_A
\]
\[
\times\left\{
\left[\frac{1}{\zeta}+\frac{1}{(1-\zeta)}-2+\zeta\bar\zeta\right]
\left(\frac{\bar \zeta}{\zeta}\right)^{2\epsilon}\left(1+\zeta^{-2\gamma}
\right)+\frac{\Gamma(1+2\epsilon)\Gamma(\frac{n}{2}-\gamma-\epsilon)
\Gamma(\frac{n}{2}+1+\gamma+\epsilon)}{\Gamma(1+\epsilon)\Gamma(1-\epsilon)
\Gamma(\frac{n}{2}-\gamma)\Gamma(\frac{n}{2}+1+\gamma+2\epsilon)}\right.
\]
\[
\times\left[\delta(1-\zeta)\left(\ln\frac{s_0}{\vec k^{\,2}}
+\psi\left(\frac{n}{2}-\gamma-\epsilon\right)
+\psi\left(1+\gamma+\frac{n}{2}+2\epsilon\right)
-\psi(\epsilon)-\psi(1)\right)+\frac{2}{(1-\zeta)_+}\right.
\]
\[
\left.\left. +\frac{\left(\zeta^{-2\epsilon-2\gamma}-1\right)}{(1-\zeta)}
+\left[\frac{1}{\zeta}-2+\zeta\bar\zeta\right]
\left(1+\zeta^{-2\epsilon-2\gamma}\right)\right]\right\}\;.
\]
Expanding this result in $\epsilon$ we obtain
\[
\frac{\alpha_s}{2\pi}\frac{\Gamma[1-\epsilon]}{\epsilon (4\pi)^\epsilon}
\frac{\Gamma^2(1+\epsilon)}{\Gamma(1+2\epsilon)}
\left(\vec k^{\,2} \right)^{\gamma+\epsilon-{n \over 2}}
\left(\vec k \cdot \vec l \,\, \right)^n
\int\limits^1_{\alpha} \frac{d\zeta}{\zeta}
f_g\left(\frac{\alpha}{ \zeta}\right)\frac{C_A}{C_F}
\]
\[
\times C_A \left\{2 \left[\frac{1}{\zeta}+\frac{1}{(1- \zeta)_+}-2
+\zeta\bar\zeta\right]\left(1+\zeta^{-2\gamma}\right)+
\delta(1-\zeta)\left(\ln\frac{s_0}{\vec k^{\,2}}+\frac{2}{\epsilon}\right)\right\}
\]
\[
=\frac{\alpha_s}{2\pi}\frac{\Gamma[1-\epsilon]}{\epsilon (4\pi)^\epsilon}
\frac{\Gamma^2(1+\epsilon)}{\Gamma(1+2\epsilon)}
\left(\vec k^{\,2} \right)^{\gamma+\epsilon-{n \over 2}}
\left(\vec k \cdot \vec l \,\, \right)^n
\int\limits^1_{\alpha} \frac{d\zeta}{\zeta}
f_g\left(\frac{\alpha}{ \zeta}\right)\frac{C_A}{C_F}
\]
\[
\times \left\{P_{gg}(\zeta)\left(1+\zeta^{-2\gamma}\right)+
\delta(1-\zeta)\left[C_A\left(\ln\frac{s_0}{\vec k^{\,2}}+\frac{2}{\epsilon}
-\frac{11}{3}\right)+\frac{2n_f}{3}\right]\right\}\;.
\]
Finally, the $\epsilon$ expansion of the divergent part has the form
\[
\left(I^R_{g;g}\right)_s=
\frac{\alpha_s}{2\pi}\frac{\Gamma[1-\epsilon]}{\epsilon(4\pi)^\epsilon}
\frac{\Gamma^2(1+\epsilon)}{\Gamma(1+2\epsilon)}
\left(\vec k^{\,2} \right)^{\gamma+\epsilon-{n \over 2}}
\left(\vec k \cdot \vec l \,\, \right)^n
\int\limits^1_{\alpha} \frac{d\zeta}{\zeta}
f_g\left(\frac{\alpha}{ \zeta}\right)\frac{C_A}{C_F}
\]
\[
\times \left\{P_{gg}(\zeta)\left(1+\zeta^{-2\gamma}\right)+
\delta(1-\zeta)\left[C_A\left(\ln\frac{s_0}{\vec k^{\,2}}
+\frac{2}{\epsilon}-\frac{11}{3}\right)+\frac{2n_f}{3}\right]\right.
\]
\[
+\, \epsilon \, C_A\left[\delta(1-\zeta)\left(\chi(n,\gamma)\ln\frac{s_0}
{\vec k^{\,2}}
+\frac{1}{2}\left(\psi^\prime\left(1+\gamma+\frac{n}{2}\right)
-\psi^\prime\left(\frac{n}{2}-\gamma\right)
-\chi^2(n,\gamma)\right)\right)\right.
\]
\[
+\left(\frac{1}{\zeta}+\frac{1}{(1-\zeta)_+}-2+\zeta\bar\zeta\right)
\left(\chi(n,\gamma)(1+\zeta^{-2\gamma})-2(1+2\zeta^{-2\gamma})\ln\zeta\right)
\]
\beq
\left.\left.
+2(1+\zeta^{-2\gamma})
\left(\left(\frac{1}{\zeta}-2+\zeta\bar\zeta\right) \ln\bar\zeta
+\left(\frac{\ln(1-\zeta)}{1-\zeta}\right)_+\right)\right]\right\}\;.
\label{GG-fin}
\eeq

For the regular part, it differs from~(\ref{regCF}) only by a factor and
reads
\[
\frac{\alpha_s}{2\pi(4\pi)^\epsilon}
\left(\vec k^{\,2} \right)^{\gamma+\epsilon-{n \over 2}}
\left(\vec k \cdot \vec l \,\, \right)^n
  \int\limits^1_{\alpha} \frac{d\zeta}{\zeta}f_{g}
\left(\frac{\alpha}{ \zeta}\right)\, \frac{C_A}{C_F}\,
\frac{\bar \zeta^2}{\zeta^2}\left[\frac{1}{\zeta}
+\frac{1}{(1- \zeta)}-2+\zeta\bar\zeta\right]
\]
\[
\times C_A\int \frac{d^{2+2\epsilon}\vec a}{\pi^{1+\epsilon}}
\frac{1}{(\vec a-\vec n)^2+\left(\vec a-\frac{\vec n}{\zeta}\right)^2}
 \left[\frac{\left(\vec a^{\,2} \right)^{\gamma-{n \over 2}}
\left(\frac{\vec a \cdot \vec l }{\vec n \cdot \vec l \,\, }\right)^n-1}
{(\vec a-\vec n)^2}+
\frac{\left(\vec a^{\,2} \right)^{\gamma-{n \over 2}}
\left(\frac{\vec a \cdot \vec l }{\vec n \cdot \vec l \,\, }\right)^n
-\zeta^{-2\gamma}}{(\vec a-\frac{\vec n}{\zeta})^2} \right]\; .
\]
Expanding it in $\epsilon$ one obtains
\beq
\left(I^R_{g;g}\right)_r=
\frac{\alpha_s}{2\pi}
\left(\vec k^{\,2} \right)^{\gamma-{n \over 2}}
\left(\vec k \cdot \vec l \,\, \right)^n \int\limits^1_{\alpha}
\frac{d\zeta}{\zeta}f_{g}\left(\frac{\alpha}{ \zeta}\right)
\frac{\bar \zeta^2}{\zeta^2}\left[\frac{1}{\zeta}+\frac{1}{(1- \zeta)}-2
+\zeta\bar\zeta\right]\frac{C_A}{C_F} C_A I_2\; .
\label{GG-reg1}
\eeq

{\bf b) both gluons generate the jet}

The jet momentum is $\vec k=\vec k_1+\vec k_2$ and the jet fraction is
$1=\zeta+\bar \zeta$. Introducing $\vec \Delta$ as
\[
\vec k_1=\zeta \vec k+\vec \Delta \;,
\]
the contribution reads
\[
I^R_{g;g+g}=
\frac{\alpha_s}{2\pi(4\pi)^\epsilon}\left(\vec k^{\,2}
\right)^{\gamma-{n \over 2}}\left(\vec k \cdot \vec l \,\, \right)^n
f_{g}\left(\alpha\right) \frac{C_A}{C_F} \int\limits^1_{0}   d\zeta \,
 \frac{C_A}{2}\left[\frac{1}{\zeta}
+\frac{1}{(1- \zeta)}-2+\zeta\bar\zeta\right]
\]
\beq
\times
\int\frac{ d^{2+2\epsilon}\vec \Delta}{\pi^{1+\epsilon}}
\left\{\frac{\vec k^{\,2}}{(\zeta \vec k+\vec \Delta)^2
(\bar \zeta \vec k-\vec \Delta)^2}+ \frac{\zeta^2 \vec k^{\,2} }
{(\zeta \vec k+\vec \Delta)^2 \vec\Delta^2}+ \frac{\bar \zeta^2 \vec k^{\,2}}
{(\bar \zeta \vec k-\vec \Delta)^2 \vec \Delta^2 }\right\}\;.
\eeq
In the SCA we need to consider only
\[
\frac{\alpha_s}{2\pi(4\pi)^\epsilon}\left(\vec k^{\,2}
\right)^{\gamma-{n \over 2}}\left(\vec k \cdot \vec l \,\, \right)^n
f_{g}\left(\alpha\right) \frac{C_A}{C_F} \int\limits^1_{0}   d\zeta \,
C_A\left[\frac{1}{\zeta}+\frac{1}{(1- \zeta)}-2+\zeta\bar\zeta\right]
\int^{\vec \Delta_{\rm max}^2}\frac{d^{2+2\epsilon}\vec \Delta}
{\pi^{1+\epsilon}}\,\frac{1}{\vec \Delta^2}\;,
\]
where $|\vec \Delta_{\rm{max}}|=|\vec k| R \min (\zeta,\bar \zeta)$. Using
\[
\int\limits^{\vec \Delta^2_{\rm{max}}}\frac{ d^{2+2\epsilon}\vec \Delta}
{\pi^{1+\epsilon}}\frac{1}{\vec \Delta^2}=\frac{1}{\epsilon\Gamma(1+\epsilon)}
(\vec\Delta^2_{\rm{max}})^\epsilon\approx\frac{\Gamma(1-\epsilon)
\Gamma^2(1+\epsilon)}{\epsilon\Gamma(1+2\epsilon)}
(\vec\Delta^2_{\rm{max}})^\epsilon \;,
\]
we get
\[
I^R_{g;g+g}=
\frac{\alpha_s}{2\pi}\frac{\Gamma(1-\epsilon)}{\epsilon (4\pi)^\epsilon}
\frac{\Gamma^2(1+\epsilon)}{\Gamma(1+2\epsilon)}
\left(\vec k^{\,2} \right)^{\gamma+\epsilon-{n \over 2}}
\left(\vec k \cdot \vec l \,\, \right)^n  f_{g}\left(\alpha\right)
\frac{C_A}{C_F} C_A R^{2\epsilon}
\]
\beq
\times\int\limits^1_{0}   d\zeta \, (\min (\zeta,\bar \zeta))^{2\epsilon}
\left[\frac{1}{\zeta}+\frac{1}{(1- \zeta)}-2+\zeta\bar\zeta\right]
\eeq
\[
\approx
\frac{\alpha_s}{2\pi}\frac{\Gamma(1-\epsilon)}{\epsilon (4\pi)^\epsilon}
\frac{\Gamma^2(1+\epsilon)}{\Gamma(1+2\epsilon)}
\left(\vec k^{\,2} \right)^{\gamma+\epsilon-{n \over 2}}
\left(\vec k \cdot \vec l \,\, \right)^n
f_{g}\left(\alpha\right)
\]
\[
\times \frac{C_A}{C_F} C_A R^{2\epsilon}
\left[\frac{1}{\epsilon}-\frac{11}{6}
+\epsilon \left(\frac{137}{36}-\frac{\pi^2}{3}+\frac{11}{3}\ln 2\right)
\right]\;.
\]

{\bf c) gluon ``inclusive'' jet generation with the other gluon in the jet
cone}

We introduce the vector $\vec \Delta$ such that
\[
\vec q=\frac{\vec k}{\zeta}+\vec \Delta\;,
\]
where $\vec k$ coincides with $\vec k_1$, the transverse momentum of the gluon
generating the jet. The contribution reads
\[
I^R_{g;g,-g}=
-\frac{\alpha_s}{2\pi(4\pi)^\epsilon}\left(\vec k^{\,2}
\right)^{\gamma-{n \over 2}} \left(\vec k \cdot \vec l \,\, \right)^n
\int\limits^1_{\alpha}   \frac{d\zeta}{\zeta} \zeta^{-2\gamma}\, f_{g}
\left(\frac{\alpha}{ \zeta}\right) \frac{C_A}{C_F} C_A
\]
\beq
\times\left[\frac{1}{\zeta}+\frac{1}{(1- \zeta)}-2+\zeta\bar\zeta\right]
\int\limits^{\vec \Delta^2_{\rm{max}}}\frac{ d^{2+2\epsilon}\vec \Delta}
{\pi^{1+\epsilon}}\frac{2}{\vec \Delta^2}\;,
\eeq
where now $|\vec\Delta_{\rm{max}}|=|\vec k| R \frac{\bar \zeta}{\zeta}$. We get
\[
I^R_{g;g,-g}=
-\frac{\alpha_s}{2\pi}\frac{\Gamma(1-\epsilon)}{\epsilon (4\pi)^\epsilon}
\frac{\Gamma^2(1+\epsilon)}{\Gamma(1+2\epsilon)}
\left(\vec k^{\,2} \right)^{\gamma+\epsilon-{n \over 2}}
\left(\vec k \cdot \vec l \,\, \right)^n R^{2\epsilon}
\]
\beq
\times\int\limits^1_{\alpha}   \frac{d\zeta}{\zeta}
\,  f_{g}\left(\frac{\alpha}{ \zeta}\right) \frac{C_A}{C_F}
C_A\left(\frac{\bar \zeta}{\zeta}\right)^{2\epsilon}
2 \zeta^{-2\gamma} \left[\frac{1}{\zeta}+\frac{1}{(1- \zeta)}
-2+\zeta\bar\zeta\right]
\eeq
\[
\approx
-\frac{\alpha_s}{2\pi}\frac{\Gamma(1-\epsilon)}{\epsilon (4\pi)^\epsilon}
\frac{\Gamma^2(1+\epsilon)}{\Gamma(1+2\epsilon)}
\left(\vec k^{\,2} \right)^{\gamma+\epsilon-{n \over 2}}
\left(\vec k \cdot \vec l \,\, \right)^n \int\limits^1_{\alpha}
\frac{d\zeta}{\zeta}\,  f_{g}\left(\frac{\alpha}{ \zeta}\right)
\frac{C_A}{C_F} C_A
\]
\[
\times
R^{2\epsilon} \left\{ \delta(1-\zeta)\frac{1}{\epsilon} + 2 \zeta^{-2\gamma}
\left[\frac{1}{\zeta}+\frac{1}{(1- \zeta)_+}-2+\zeta\bar\zeta\right]\right.
\]
\[
\left.
+ \,4\, \epsilon\, \zeta^{-2\gamma}\left(\left(\frac{1}{\zeta}-2
+\zeta\bar\zeta\right)\ln\frac{\bar \zeta}{\zeta} -\frac{\ln \zeta}{1-\zeta}
+\left(\frac{\ln (1-\zeta)}{1-\zeta}\right)_+\right)\right\}\;.
\]
Introducing the splitting function $P_{gg}$, we get
\[
I^R_{g;g,-g}=
-\frac{\alpha_s}{2\pi}\frac{\Gamma(1-\epsilon)}{\epsilon (4\pi)^\epsilon}
\frac{\Gamma^2(1+\epsilon)}{\Gamma(1+2\epsilon)}
\left(\vec k^{\,2} \right)^{\gamma+\epsilon-{n \over 2}}
\left(\vec k \cdot \vec l \,\, \right)^n \int\limits^1_{\alpha}
\frac{d\zeta}{\zeta}\, f_{g}\left(\frac{\alpha}{ \zeta}\right) \frac{C_A}{C_F}
\]
\[
\times
R^{2\epsilon} \left\{ P_{gg}(\zeta)\zeta^{-2\gamma} +\delta(1-\zeta)
\left[ C_A\left(\frac{1}{\epsilon}-\frac{11}{6}\right)+\frac{n_f}{3}\right]
\right.
\]
\[
\left.
+ \, \epsilon\,4 C_A\,  \zeta^{-2\gamma}\left(\left(\frac{1}{\zeta}
-2+\zeta\bar\zeta\right)\ln\frac{\bar \zeta}{\zeta} -\frac{\ln \zeta}
{1-\zeta}+\left(\frac{\ln (1-\zeta)}{1-\zeta}\right)_+\right)\right\}\;.
\]

\subsubsection{Final result for the case of incoming gluon}

Collecting all the contributions calculated in this Section and taking
into account the PDFs' renormalization counterterm~(\ref{c.count.t})
and the charge counterterm~(\ref{charge.count.t}), we find that all singular
contributions cancel and the result is
\beq
I_g=\frac{\alpha_s}{2\pi}\left(\vec k^{\,2} \right)^{\gamma-{n \over 2}}
\left(\vec k \cdot \vec l \,\, \right)^n \int\limits^1_{\alpha}
\frac{d\zeta}{\zeta}f_{g}\left(\frac{\alpha}{ \zeta}\right)\frac{C_A}{C_F}
\label{resultg}
\eeq
\[
\times
\left\{
\left\{P_{gg}(\zeta)+2 \,n_f \frac{C_F}{C_A}P_{qg}(\zeta)\right\}
\ln\frac{\vec k^{\,2}}{\mu_F^2}-
2\zeta^{-2\gamma}\ln R \left\{P_{gg}(\zeta)+2 \,n_f P_{qg}(\zeta)\right\}
-\frac{\beta_0}{2}\ln\frac{\vec k^{\,2}}{4\mu_R^2}\delta(1-\zeta)
\right.
\]
\[
+\, C_A\delta(1-\zeta)
\left\{
\chi(n,\gamma)\ln\frac{s_0}{\vec k^{\,2}}+\frac{1}{12}+\frac{\pi^2}{6}
+\frac{1}{2}\left(\psi^\prime\left(1+\gamma+\frac{n}{2}\right)
-\psi^\prime\left(\frac{n}{2}-\gamma\right)-\chi^2(n,\gamma)\right)
\right\}
\]
\[
+\, 2 C_A (1-\zeta^{-2\gamma})\left(\left(\frac{1}{\zeta}-2
+\zeta\bar\zeta\right)\ln \bar \zeta + \frac{\ln (1-\zeta)}{1-\zeta}
\right)
\]
\[
+ \, C_A\, \left[\frac{1}{\zeta}+\frac{1}{(1- \zeta)_+}-2+\zeta\bar\zeta\right]
\left((1+\zeta^{-2\gamma})\chi(n,\gamma)-2\ln\zeta+\frac{\bar \zeta^2}
{\zeta^2}I_2\right)
\]
\[
\left.
+\, n_f\left[\, 2\zeta\bar \zeta \, \frac{C_F}{C_A} +(\zeta^2+\bar \zeta^2)
\left(\frac{C_F}{C_A}\chi(n,\gamma)+\frac{\bar \zeta}{\zeta}I_3\right)
-\frac{1}{12}\delta(1-\zeta)\right]\right\}\;.
\]

For the $I_{1,2,3}$ functions, which enter our final expressions for the quark
and gluon contributions, we obtain the following results:
\beq
I_2=\frac{\zeta^2}{\bar \zeta^2}\left[
\zeta\left(\frac{{}_2F_1(1,1+\gamma-\frac{n}{2},2+\gamma-\frac{n}{2},\zeta)}
{\frac{n}{2}-\gamma-1}-
\frac{{}_2F_1(1,1+\gamma+\frac{n}{2},2+\gamma+\frac{n}{2},\zeta)}{\frac{n}{2}+
\gamma+1}\right)\right.
\eeq
$$
\left.
+\zeta^{-2\gamma}
\left(\frac{{}_2F_1(1,-\gamma-\frac{n}{2},1-\gamma-\frac{n}{2},\zeta)}
{\frac{n}{2}+\gamma}-
\frac{{}_2F_1(1,-\gamma+\frac{n}{2},1-\gamma+\frac{n}{2},\zeta)}{\frac{n}{2}
-\gamma}\right)
\right.
$$
$$
\left.
+\left(1+\zeta^{-2\gamma}\right)\left(\chi(n,\gamma)-2\ln \bar \zeta \right)
+2\ln{\zeta}\right]\;,
$$

\beq
I_1=\frac{\bar \zeta}{2\zeta}I_2+\frac{\zeta}{\bar \zeta}\left[
\ln \zeta+\frac{1-\zeta^{-2\gamma}}{2}\left(\chi(n,\gamma)-2\ln \bar \zeta
\right)\right]\;,
\eeq

\beq
I_3=\frac{\bar \zeta}{2\zeta}I_2-\frac{\zeta}{\bar \zeta}\left[
\ln \zeta+\frac{1-\zeta^{-2\gamma}}{2}\left(\chi(n,\gamma)-2\ln \bar \zeta
\right)\right]\;.
\eeq

Using the following property of the hypergeometric function,
$$
{}_2F_1(1,a,a+1,\zeta)=a\sum^\infty_{n=0}\frac{(a)_n}{n!}\left[\psi(n+1)-\psi(a+n)-\ln\bar \zeta\right]\, \bar \zeta^n\;,
$$
one can easily see that for $\zeta\to 1$,
$$
I_2={\cal O}\left(\ln \bar \zeta \right)\, ,
\quad I_1={\cal O}(\ln \bar \zeta ) \, ,\quad I_3={\cal O}(\ln \bar \zeta )\,,
$$
which implies that the integral over $\zeta$ in~(\ref{resultq})
and in~(\ref{resultg}) is convergent on the upper limit.

\section{Summary}

In this paper we have calculated the NLO vertex (impact factor) for the forward
production of high-$p_T$ jet from an incoming quark or gluon, emitted
by a proton, in the ``small-cone'' approximation. This vertex is an ingredient
for the calculation of the hard inclusive production of a pair of forward
high-$p_T$ (or Mueller-Navelet) jets in proton collisions.

At the basis of the calculation of the hard part of the vertex was the
definition of NLO BFKL parton impact factors; then the collinear factorization
(in the $\overline{\rm{MS}}$ scheme) with the PDFs of the incoming partons
was suitably considered.

We have presented our result for the vertex in the so called
$(\nu,n)$-representation, which is the most convenient one in view of the
numerical determination of the cross section for the production of
a pair of rapidity-separated jets, along the same lines as in
Ref.~\cite{mesons}.

We have explicitly verified that soft and virtual infrared divergences cancel
each other, whereas the infrared collinear ones are compensated by the PDFs'
renormalization counterterms, the remaining ultraviolet divergences
being taken care of by the renormalization of the QCD coupling.

In our approach the energy scale $s_0$ is an arbitrary parameter, that need not
be fixed at any definite scale. The dependence on $s_0$ will disappear
in the next-to-leading logarithmic approximation in any physical cross
section in which jet vertices are used. Indeed,  our result for the NLO jet
vertex, given by Eqs.~(\ref{def_I}),~(\ref{def_I_q+g}),~(\ref{resultq})
and~(\ref{resultg}) contains  contributions $\sim \ln(s_0)$ and these terms
are proportional to the LO quark and gluon jet vertices multiplied by the BFKL
kernel  eigenvalue  $\chi(n,\nu)$. This fact guarantees the independence
of the jet cross
section on $s_0$ within the next-to-leading logarithmic approximation.
However, the dependence on this energy scale will survive in terms beyond
this approximation and will provide a parameter to be optimized with the
method adopted in Refs.~\cite{mesons}.

The small-cone approximation, which we adopted here, allows us to obtain
explicit analytical result for the jet impact factor.
In the general case the dependence of the partonic cross section on the jet
cone parameter has, in the limit $R\to 0$, the form $d\sigma\sim
A \ln R+B+{\cal O}(R^2)$ (see, for instance,~\cite{Furman:1981kf} and
Appendix C there). In fact, in our work we calculated the coefficients $A$ and
$B$, neglecting all pieces ${\cal O}(R^2)$. This can be seen directly from our
formulas; for example in proceeding from  Eq.~(\ref{example1}) to
Eq.~(\ref{example2})  the contributions ${\cal O}(\Delta^2)\sim
{\cal O}(R^2)$ were neglected.
The quality of the small-cone approximation has been checked by comparison
with the results of Monte Carlo calculations which treat the cone size exactly,
both for the cases of unpolarized and polarized jet cross sections.
Very good agreement between the results of the small-cone approximation and
the Monte Carlo calculations was found even for cone sizes of up to $R=0.7$,
 see~\cite{Jager:2004jh} for more details and references.
Therefore having this experience with the jet production in Bjorken kinematics
$s\sim Q^2$, there is the hope that the small-cone approximation could also be
an adequate tool for describing Mueller-Navelet jets for an experimentally
relevant choice of the jet cone parameter, $R\sim 0.5$.
Another important application of the small-cone approximation method could be
the possibility to perform a semi-analytical check of the complicated
numerical approaches to Mueller-Navelet jet production which treat the cone
size exactly,  in the limit of small values of $R$.

\section*{Acknowledgements}

We thank A. Kotikov for a useful discussion.
D.I. thanks the Dipartimento di Fisica dell'U\-ni\-ver\-si\-t\`a della Calabria
and the Istituto Nazio\-na\-le di Fisica Nucleare (INFN), Gruppo collegato di
Cosenza, for the warm hospitality and the financial support. This work was
also supported in part by the grants and RFBR-11-02-00242 and NSh-3810.2010.2.

\appendix
\section{Appendix}

We list here some useful integrals
\beq
\int
\frac{d^{2+2\epsilon}\vec k^\prime}{\vec k^{\prime \,2}}
\left(\frac{1}{\vec k^{\prime \,2}+(\vec k^\prime-\vec k \,)^2}\right) =
\frac{1}{2}\int
\frac{d^{2+2\epsilon}\vec k^\prime}{\vec k^{\prime\,2}(\vec k^\prime-\vec k)^2}
=\pi^{1+\epsilon}\left(\vec k^{\,2}\right)^{\epsilon -1}
\frac{\Gamma(1-\epsilon) \Gamma^2(1+\epsilon)}
{\epsilon\, \Gamma(1+2\epsilon)}\;,
\label{I0}
\eeq

\beq
\int \frac{d^{2+2\epsilon}\vec k^\prime (\vec k^{\prime\,2})^\alpha}
{(\vec k-\vec k^\prime)^2} =
\pi^{1+\epsilon}\left(\vec k^{\,2}\right)^{\alpha + \epsilon}
\frac{\Gamma(-\epsilon-\alpha)}{\Gamma(-\alpha)}\frac{\Gamma(\epsilon)\,
\Gamma(1+\epsilon+\alpha)}{\Gamma(1+\alpha+2\epsilon)}\;.
\eeq

In the integrals below, $\vec l^{\;2}=0$ is assumed

\beq
\int \frac{\displaystyle d^{2+2\epsilon} \vec k^\prime
(\vec k^{\prime\,2})^\alpha (\vec k^\prime \cdot \vec l \, )^\beta}
{\displaystyle (\vec k-\vec k^\prime)^2} =
\pi^{1+\epsilon}  \left( \vec k \cdot \vec l\, \right)^{\beta}
\left(\vec k^{\,2}\right)^{\alpha +\epsilon}
\frac{\displaystyle \Gamma(-\alpha-\epsilon)}{\displaystyle \Gamma(-\alpha)}
\frac{\displaystyle \Gamma(\epsilon)\, \Gamma(1+\epsilon+\alpha+\beta)}
{\displaystyle \Gamma(1+\alpha+\beta+2\epsilon)}\;,
\label{I2}
\eeq

\beq
{\displaystyle \int} \frac{\displaystyle d^{2+2\epsilon} \vec k^\prime
\ln (\vec k-\vec k^\prime)^2  (\vec k^{\prime\,2})^\alpha
(\vec k^\prime \cdot \vec l\, )^\beta}{\displaystyle (\vec k-\vec k^\prime)^2}
=\pi^{1+\epsilon}  \left( \vec k \cdot \vec l\, \right)^{\beta}
\left(\vec k^{\,2}\right)^{\alpha +\epsilon}
\eeq
\[
\times\frac{\displaystyle \Gamma(-\alpha-\epsilon)}{\displaystyle
\Gamma(-\alpha)}\frac{\displaystyle \Gamma(\epsilon)\,
\Gamma(1+\epsilon+\alpha+\beta)}
{\displaystyle \Gamma(1+\alpha+\beta+2\epsilon)}
\]
\[
\times \left\{\ln \vec k^{\,2} +\psi(\epsilon)+ \psi(1)- \psi(-\alpha-\epsilon)
-\psi(1+\alpha+\beta+2\epsilon) \right\}\;,
\]

\beq
\int\limits^{2\pi}_0 d\phi\frac{\cos n\phi}{a^2-2 a b \cos\phi +b^2 }
=\frac{2\pi}{b^2-a^2}\left(\frac{a}{b}\right)^n\ , \quad a<b \, ,
\quad n \geq 0\;.
\eeq

\end{document}